\documentclass[acmsmall]{acmart}

\usepackage{makecell}
\usepackage{multirow,multicol}
\usepackage{subcaption}
\usepackage{tcolorbox}
\usepackage{threeparttable}
\usepackage{xspace}
\usepackage{enumitem}
\usepackage{graphicx}
\usepackage{multirow}
\usepackage{xcolor}
\usepackage[table]{xcolor}
\usepackage{tabularx}
\usepackage[normalem]{ulem}

\long\def\add#1{#1}
\long\def\delete#1{}
\long\def\delpar#1{}

\usepackage[table]{xcolor}
\usepackage{colortbl}

\newcommand{\modelname}{GrepRAG\xspace}
\newcommand{\stitle}[1]{\vspace{0.1mm} \noindent {\bf #1}}

\AtBeginDocument{%
  }

\setcopyright{cc}
\setcctype{by}
\acmDOI{10.1145/3832136}
\acmYear{2026}
\acmJournal{PACMSE}
\acmVolume{3}
\acmNumber{ISSTA}
\acmArticle{ISSTA045}
\acmMonth{10}
\acmSubmissionID{issta26main-p388-p}
\received{2026-01-30}
\received[accepted]{2026-06-25}

\begin{document}

\title{Better Call Grep: Evaluating and Improving Grep-Like Lexical Retrieval for Repository-Level Code Completion}


\author{Baoyi Wang}
\orcid{0009-0002-3123-2982}
\authornote{Both authors contributed equally to this research.}
\affiliation{%
  \institution{Zhejiang University}
  \city{Ningbo}
  \country{China}
}
\email{wangbaoyi@zju.edu.cn}

\author{Xingliang Wang}
\orcid{0000-0003-2843-4831}
\authornotemark[1]
\affiliation{%
  \institution{Zhejiang University}
  \city{Hangzhou}
  \country{China}
}
\email{wangxingliang@zju.edu.cn}

\author{Guochang Li}
\orcid{0009-0000-0770-9133}
\affiliation{%
  \institution{Zhejiang University}
  \city{Hangzhou}
  \country{China}
}
\email{gcli@zju.edu.cn}

\author{Chen Zhi}
\orcid{0000-0002-1273-2992}
\authornote{Corresponding authors: Chen Zhi and Junxiao Han.}
\affiliation{%
  \institution{Zhejiang University}
  \city{Ningbo}
  \country{China}
}
\email{zjuzhichen@zju.edu.cn}

\author{Junxiao Han}
\orcid{0000-0001-7630-8667}
\authornotemark[2]
\affiliation{%
  \institution{Hangzhou City University}
  \city{Hangzhou}
  \country{China}
}
\email{hanjx@hzcu.edu.cn}

\author{Xinkui Zhao}
\orcid{0000-0002-1115-5652}
\affiliation{%
  \institution{Zhejiang University}
  \city{Hangzhou}
  \country{China}
}
\email{zhaoxinkui@zju.edu.cn}

\author{Nan Wang}
\orcid{0009-0003-3410-078X}
\affiliation{%
  \institution{Shenzhou Aerospace Software Technology}
  \city{Beijing}
  \country{China}
}
\email{wangnan02026@163.com}

\author{Shuiguang Deng}
\orcid{0000-0001-5015-6095}
\affiliation{%
  \institution{Zhejiang University}
  \city{Hangzhou}
  \country{China}
}
\email{dengsg@zju.edu.cn}

\author{Jianwei Yin}
\orcid{0000-0003-4703-7348}
\affiliation{%
  \institution{Zhejiang University}
  \city{Hangzhou}
  \country{China}
}
\email{zjuyjw@cs.zju.edu.cn}

\renewcommand{\shortauthors}{B. Wang, X. Wang, G. Li, C. Zhi, J. Han, X. Zhao, N. Wang, S. Deng, and J. Yin}
\title[Better Call Grep: Grep-Like Retrieval for Repository-Level Code Completion]{Better Call Grep: Evaluating and Improving Grep-Like Lexical Retrieval for Repository-Level Code Completion}

\begin{abstract}
  
 Repository-level code completion remains challenging for large language models (LLMs), as it requires reasoning over cross-file dependencies while under limited context windows. To address this challenge, prior work has adopted Retrieval-Augmented Generation (RAG) frameworks based on semantic indexing or structure-aware graph analysis. Although effective, these approaches introduce substantial computational overhead for index construction and maintenance, which hinders their practicality in real-world development. Motivated by common developer workflows that rely on lightweight search utilities (e.g., \texttt{ripgrep}) to locate relevant code, we revisit a fundamental yet underexplored question: how far can simple, index-free lexical retrieval go in supporting repository-level code completion before more complex retrieval mechanisms become necessary? To answer this question, we systematically explore the potential of lightweight, index-free, intent-aware lexical retrieval through extensive empirical analysis. We first introduce Naive \modelname, a baseline framework where LLMs autonomously generate ripgrep commands to localize relevant context. Our preliminary experiments show that even this basic implementation achieves performance comparable to sophisticated graph-based baselines. Further analysis reveals that its effectiveness stems from retrieving code fragments that are lexically precise and spatially closer to the completion site. However, we identify key limitations of this approach, including sensitivity to noisy matches caused by high-frequency ambiguous keywords and context fragmentation due to rigid truncation boundaries. To address these issues, we propose \modelname, which augments lexical retrieval with a lightweight post-processing pipeline featuring identifier-weighted re-ranking and structure-aware deduplication. Extensive evaluation on CrossCodeEval and RepoEval\_Updated demonstrates that \modelname\ consistently outperforms state-of-the-art (SOTA) methods. In particular, on CrossCodeEval, \modelname\ achieves 7.04–15.58\% relative improvement in code exact match (EM) over the best baseline.
\end{abstract}





\maketitle

\section{Introduction}

As a core feature of modern Integrated Development Environments (IDEs), code completion plays a pivotal role in enhancing development efficiency~\cite{hindle2016naturalness,amann2016study,raychev2014code,ziegler2022productivity,sun2025don,zhao2025completion}. In recent years, automated code completion driven by LLMs has demonstrated remarkable proficiency within single-file or single-function contexts~\cite{pan2025code,zan2022large,chen2021evaluating,roziere2023code,han2021empirical,zhi2024llm}. However, its performance often degrades significantly when applied to repository-level code completion~\cite{liu2023repobench,wang2025teaching,vasconcelos2025generation,li2025deepcircuitx,li2025empowering}. In large-scale repositories, relevant contextual information is typically fragmented across multiple files and directories due to modular code organization. As a result, crucial dependencies such as class definitions, utility interfaces, and global constants are frequently located outside the local file being edited. Given the limited context window of LLMs, it is infeasible to provide the model with the entire repository. Relying solely on intra-file context therefore misses essential cross-file dependencies~\cite{ding2023crosscodeeval,shi2023large,liu2023lost}.

RAG mitigates this contextual deficit by retrieving code snippets most relevant to the current logic at the repository level~\cite{rag,yang2025empirical,parvez2021retrieval}. Existing RAG methodologies can be primarily categorized into three streams. The first category comprises traditional similarity-based retrieval methods~\cite{zhang2023repocoder,shrivastava2023repofusion,shrivastava2023repository}, which rank code snippets directly based on cosine similarity~\cite{reimers2019sentence} or BM25~\cite{robertson2009probabilistic} scores relative to the code completion context. The second category involves structure-aware retrieval~\cite{liu2024graphcoder,liang2024repofuse,phan2024repohyper,ding2022cocomic,wang2025grace}, which models the repository as a dependency graph via static analysis and leverages graph structural information to locate relevant context. The third category pertains to strategy-optimized retrieval~\cite{wang2024rlcoder}; these methods employ Reinforcement Learning (RL) to train dense retrievers, thereby dynamically optimizing retrieval strategies based on end-to-end completion feedback.

Although existing RAG methods are effective, they often rely on complex preprocessing and index construction, which impose substantial time and computational costs. For example, on the \texttt{huggingface\_diffusers} repository\footnote{\url{https://github.com/huggingface/diffusers}} , which contains about 100K lines of Python code, GraphCoder~\cite{liu2024graphcoder} requires approximately 91 seconds to build the graph index and 7 seconds for retrieval. In contrast, developers typically expect latency below 0.5 seconds, while delays exceeding 2 seconds are considered unacceptable for user experience~\cite{wang2023practitioners}. Furthermore, software repositories are dynamic, frequent code modifications render static graphs and vector indices stale. 

Inspired by human developers, who commonly use simple lexical search tools such as \texttt{Grep}, \texttt{Ctrl+F}, or IDE features like \textit{Go to Definition} and \textit{Go to Implementation} during programming to locate class and method definitions or implementations across files.
Such tools are also used to retrieve code fragments with similar naming patterns distributed throughout the repository. This structural or lexically similar information is essential for resolving cross-file dependencies in repository-level code completion.~\cite{liang2024repofuse} The observation motivates us to reconsider whether the potential of simple lexical retrieval has been fully explored before resorting to complex structural or semantic retrieval mechanisms. In parallel, modern agentic coding systems such as ClaudeCode and Windsurf have preliminarily demonstrated the feasibility of simple lexical matching by integrating Grep tools for multi-turn code question-answering tasks; however, the systematic application and evaluation of such techniques within the specific context of {code completion} remain unexplored.

\begin{figure*}[tbp] 
\centering \includegraphics[width=\textwidth]{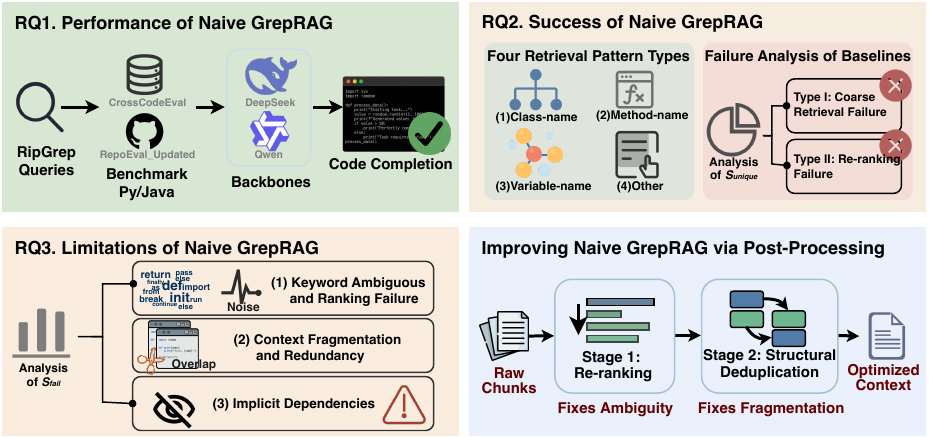} 
\caption{Overview of the research framework. We first evaluate the effectiveness of Naive \modelname\ (RQ1), analyze the factors contributing to its success (RQ2), identify its limitations (RQ3), and finally propose an optimized \modelname\ equipped with a post-processing pipeline.}
\label{fig:introduction} 
\end{figure*}

To systematically study this problem, we adopt a progressive research that first evaluates the potential of grep-like lexical retrieval for repository-level code completion and then explores methods to improve it. The overall framework is illustrated in Figure~\ref{fig:introduction}.
First, we introduce Naive \modelname, a framework where the LLM autonomously generates \texttt{ripgrep} commands for context localization, establishing a performance benchmark for lexical retrieval in repository-level code completion. 
Our exploratory experiments show that the performance of this naive approach outperforms complex graph-based methods (RQ1). 
To further understand why Naive \modelname performs effectively, we analyze the retrieval patterns of Naive \modelname and identify four main types of keywords, including class names, method names, variable names, and others, which are particularly effective for handling several completion scenarios, such as class declaration completion, method call completion, etc. Compared with baseline failures, Naive \modelname succeeds by retrieving code fragments closer to the completion site with more precise lexical queries (RQ2).
However, our analysis in RQ1 reveals that Naive \modelname does not cover all cases solved by baseline methods, prompting a investigation of its limitations. Our analysis identifies two main issues: (1) {keyword ambiguity}, where high-frequency generic terms such as \texttt{init} introduce noise; and (2) {context redundancy and fragmentation}, where overlapping code blocks are truncated without merging. This not only results in redundant information occupying the context window, but also disrupts the integrity of the code flow due to fragmentation (RQ3).
Based on these empirical insights, we propose \modelname. This enhanced framework introduces a lightweight post-processing pipeline incorporating identifier-weighted re-ranking and structure-aware deduplication. This approach addresses the lack of term weighting in the Jaccard algorithm. It also mitigates the issues of context fragmentation and redundancy caused by Grep-style retrieval. By resolving these problems, \modelname significantly improves completion performance while maintaining efficient retrieval.

In summary, the primary contributions of this paper are outlined as follows:

\begin{itemize}
    \item We evaluate the potential of Naive \modelname and demonstrate that even a lightweight, grep-based RAG framework can achieve competitive performance on benchmark datasets.
    
    \item We analyze the factors behind Naive \modelname's success, examining its retrieval patterns and contrasting them with failures of RAG-based baselines, providing insights for future retrieval-augmented code completion research.
    
    \item We delineate the limitations of Naive \modelname and introduce an optimized \modelname strategy that incorporates identifier-weighted re-ranking and structure-aware deduplication. This approach tackles keyword ambiguity, context redundancy, and fragmentation, achieving SOTA performance across multiple benchmarks.

\end{itemize}
\section{Motivation}

\subsection{Quantitative Analysis of Retrieval Latency and Scalability}

While existing Vanilla-RAG and GraphRAG~\cite{liu2024graphcoder} approaches demonstrate superior performance in enhancing completion accuracy, their retrieval latency can be substantial in practice. To systematically assess this cost, we measured the retrieval latency of BM25-based Vanilla-RAG and GraphCoder on the repositories of the RepoEval\_Updated~\cite{liu2024graphcoder} dataset.

\begin{table}[htbp]
\centering
\caption{Average Retrieval Latency (seconds) on RepoEval\_Updated Dataset. The time represents the average inference latency of Line and API completion tasks. \textbf{Bold} indicates latency exceeding 2 seconds, which is considered unacceptable for real-time completion.}
\label{tab:repoeval_latency}
\resizebox{\textwidth}{!}{%
\begin{tabular}{lrrr|lrrr}
\toprule
\multicolumn{4}{c|}{\textbf{Python Repositories}} & \multicolumn{4}{c}{\textbf{Java Repositories}} \\
\midrule
\textbf{Repository} & \textbf{LOC(K)} & \textbf{VanillaRAG (s)} & \textbf{GraphCoder (s)}
 & \textbf{Repository} & \textbf{LOC(K)} & \textbf{VanillaRAG (s)} & \textbf{GraphCoder (s)}
 \\ 
\midrule
devchat & 3.2 & 0.036 & 0.307 & chatgpt4j & 5.2 & 0.102 & 0.249 \\
nemo\_aligner & 6.8 & 0.122 & 0.672 & Harmonic-HN & 10.5 & 0.530 & 0.922 \\
task\_weaver & 10.9 & 0.137 & 0.988 & rusty-connector & 11.4 & 0.263 & 0.622 \\
awslabs\_fortuna & 16.5 & 0.237 & 0.965 & NeoGradle & 14.4 & 0.585 & 1.139 \\
nerfstudio & 22.7 & 0.484 & \textbf{2.056} & open-dbt & 26.9 & 0.657 & 1.198 \\
metagpt & 27.1 & 0.297 & \textbf{2.357} & mybatis-flex & 64.3 & 1.688 & \textbf{3.400} \\
opendilab\_ACE & 59.1 & 0.877 & \textbf{5.058} & rocketmq & 120.3 & \textbf{3.040} & \textbf{7.067} \\
diffusers & 82.1 & \textbf{3.017} & \textbf{6.900} & pixel-dungeon & 147.7 & \textbf{5.023} & \textbf{10.836} \\
apple\_axlearn & 132.3 & \textbf{6.615} & \textbf{9.395} & cms-oss & 493.5 & \textbf{65.051} & \textbf{28.400} \\
AdaLoRA & 577.8 & \textbf{97.265} & \textbf{46.765} & FloatingPoint & 753.9 & \textbf{53.822} & \textbf{50.463} \\
\bottomrule
\end{tabular}%
}
\end{table}

As shown in Table~\ref{tab:repoeval_latency}, existing retrieval methods incur substantial latency when applied to large-scale repositories (e.g., \texttt{AdaLoRA} and \texttt{FloatingPoint}, each exceeding 500K lines of code). Specifically, a single retrieval using \texttt{VanillaRAG} or \texttt{GraphCoder} can exceed 40 seconds.

It is important to note that Table~\ref{tab:repoeval_latency} reports only the retrieval latency, excluding preprocessing time such as index construction and static graph modeling. This significant temporal overhead prompts a critical reassessment of existing approaches: Is such a computational cost truly unavoidable for acquiring the context necessary for code completion?

Retrieval tools based on lexical retrieval offer an efficient, index-free alternative to the aforementioned latency bottleneck. We conducted preliminary tests using ripgrep, performing retrieval via lexical matching within the RAG framework, thereby avoiding the overhead of index and graph construction. As shown in Table~\ref{tab:repoeval_latency} for the \texttt{diffusers} repository, retrieval via \texttt{ripgrep} required only approximately {0.40s}, whereas baseline methods consumed between 3s and 7s. This performance disparity becomes even more pronounced in large-scale repositories. For instance, within the \texttt{FloatingPoint} Java repository (containing 754k LOC), the retrieval overhead of existing methods exceeded 50s. In contrast, \texttt{grep}-based retrieval required only {1.45s}, representing a reduction to approximately 1/35th of the original time cost. These results suggest that lexical retrieval, when coupled with its zero-indexing requirement, offers a compelling trade-off between efficiency and practicality for large-scale repository-level code completion.

\subsection{Qualitative Analysis of Retrieval Effectiveness}

\begin{figure}[t]
    \centering
    \includegraphics[width=0.98\textwidth]{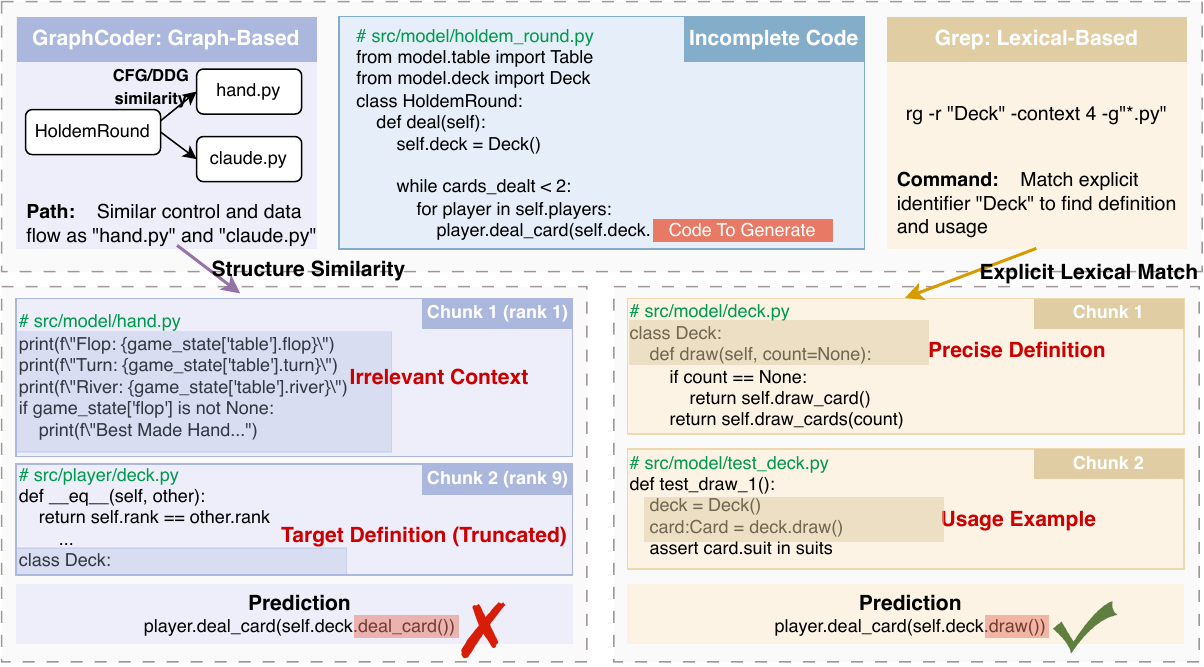}
    \caption{Comparison of GraphCoder and Grep in a method invocation scenario. \textbf{Left:} GraphCoder retrieves irrelevant chunks. \textbf{Right:} Grep locates the precise definition and usage examples via lexical retrieval.}
    \label{fig:case_study}
    \vspace{-0.3cm}
\end{figure}

While lexical retrieval demonstrates significant efficiency advantages, its ability to capture cross-file dependencies remains uncertain due to the lack of deep semantic reasoning. To investigate this, we conducted a case study using the \texttt{TexasHoldemAgents} repository from the CrossCodeEval~\cite{ding2023crosscodeeval} dataset. Figure~\ref{fig:case_study} illustrates a representative completion scenario. In the \texttt{src/model/\allowbreak holdem\_round.py} file, a developer attempts to invoke the \texttt{draw()} method of the \texttt{self.deck} object within the \texttt{deal} function. Since the \texttt{Deck} class is defined in a separate file, \texttt{src/model/deck.py}, the model must access the precise class definition or method signature to generate an accurate prediction.

We contrast the retrieval behaviors of GraphCoder and \texttt{ripgrep} in this scenario. 
As shown in Figure~\ref{fig:case_study} (Left), GraphCoder ranks a snippet from \texttt{src/players/claude.py} as its most relevant result (rank~1), which contains logging logic and downstream method invocations unrelated to the definition of \texttt{Deck}. However, the actual class definition of \texttt{Deck} in \texttt{src/model/deck.py} is relegated to a much lower rank (rank~9) and retrieved in an incomplete, truncated form.
These retrieved results contain noisy and fragmented code blocks that offer little assistance for code completion.

In contrast, lexical retrieval based on explicit identifiers directly located concrete code definitions and usages across the repository. As illustrated in Figure~\ref{fig:case_study} (Right), the Grep command targeted the identifier \texttt{"Deck"} within the completion context, successfully recalling both class definitions and object instantiation code. This combination of definition and usage, facilitated by the lexical retrieval mechanism, provided the model with precise cross-file contextual information, effectively compensating for the absence of semantic reasoning capabilities.

\section{Experimental Setup}

Based on the aforementioned observations, we design a comprehensive experimental study to systematically evaluate the effectiveness and efficiency of grep\_like lexical retrieval for repository-level code completion.
This section introduces the experimental setup, including datasets, evaluation metrics, Naive \modelname, baseline methods, and implementation details.

\subsection{Datasets}
We evaluate on CrossCodeEval\cite{ding2023crosscodeeval} and RepoEval\_Updated\cite{liu2024graphcoder}, with statistics summarized in Table~\ref{tab:dataset-stats-updated}.
\begin{table}[htbp]
    \centering
    \caption{Statistics of the CrossCodeEval and RepoEval\_Updated Dataset Subsets}
    \label{tab:dataset-stats-updated}
    \resizebox{0.8\linewidth}{!}{%
    \begin{tabular}{lccccc}
        \toprule
        \textbf{Dataset} & \textbf{Subset} & \textbf{\#Repositories} & \textbf{\#Files} & \textbf{\#Examples} & \textbf{Avg.\#Tokens} \\
        \midrule
        \multirow{2}{*}{CrossCodeEval} & Python & 471 & 1368 & 2665 & 14.45 \\
                                        & Java   & 239 & 745  & 2139 & 16.76 \\
        \midrule
        \multirow{2}{*}{RepoEval\_Updated} & Python & 10 & 3258 & 4000 & 15.44 \\
                                         & Java   & 10 & 8522  & 4000 & 17.82 \\
        \bottomrule
    \end{tabular}
    }
\end{table}

\textbf{CrossCodeEval.}
We use the Python and Java subsets of CrossCodeEval to evaluate the model's ability to handle scenarios that strictly require cross-file context for accurate code completion. The dataset includes 471 Python and 239 Java repositories, and applies static analysis to exclude samples that can be resolved using only intra-file context. This rigorous filtering ensures that all test cases depend on external context, allowing for a precise assessment of the retrieval module's effectiveness in locating cross-file information.

\textbf{RepoEval\_Updated.}
To evaluate model performance on large-scale repositories, we employ the RepoEval\_Updated dataset, derived from RepoEval~\cite{zhang2023repocoder}. This dataset incorporates a task classification mechanism, categorizing tasks into {Line-level} for general coding and {API-level} for scenarios necessitating intra-repository API invocations. Comprising a total of 8,000 tasks across Python and Java, the dataset encompasses projects with substantial code volume (with some exceeding 500k LOC), thereby serving as an effective stress test for retrieval latency and scalability.

\subsection{Evaluation Metrics}
Following CrossCodeEval\cite{ding2023crosscodeeval}
, we assess generation quality across two dimensions: (1) {code match}, which evaluates overall textual consistency, and (2) {identifier match}, which focuses on the semantic accuracy of API calls and variables extracted via static analysis. For both dimensions, we employ four metrics:
\textbf{EM} measures strict equality between the generated sequence and reference. \textbf{Edit Similarity (ES)} quantifies similarity based on Levenshtein distance\cite{Levenshtein1965BinaryCC} ($Lev$). Additionally, \textbf{Recall} and \textbf{F1} evaluate content coverage. Notably, for identifier match, identifiers are treated as a {set} to assess prediction accuracy regardless of order, whereas code match treats code as token sequences.
We also report retrieval latency, measured as the average CPU time per query, to evaluate the efficiency of the retrieval process.

\subsection{Naive \modelname}

To investigate the performance of lexical retrieval in repository-level code completion, this study constructs Naive \modelname, as illustrated in Figure~\ref{fig:naive_model}. 
This framework consists of three phases:

\begin{figure*}[tbp] 
\centering \includegraphics[width=\textwidth]{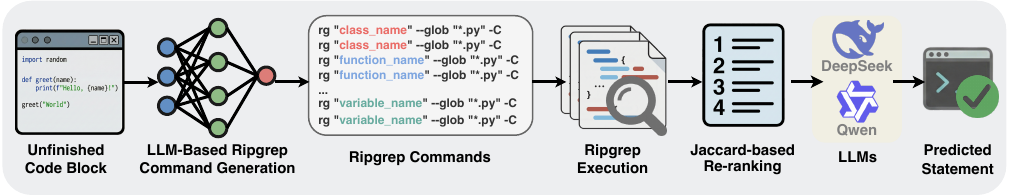} 
\caption{Overview of the Naive GrepRAG framework.}
\label{fig:naive_model} 
\vspace{-0.2cm}
\end{figure*}

\stitle{Grep Query Generation:} Given the local context preceding the cursor, $\mathcal{C}_{local}$, the LLM autonomously generates $m$ \texttt{ripgrep} commands, $\mathcal{Q} = \{q_1, q_2, \dots, q_m\}$, by analyzing the code's lexical features, latent dependencies, and the user's coding intent. In our setting, $m$ is specified as 10 via the prompt, though the actual number of generated commands may vary slightly due to the LLM’s generation behavior. The prompt is provided in the described in Section~\ref{sec:data_availability}.

\stitle{Deterministic Execution:} The query set, $\mathcal{Q}$, is executed across the repository using the \texttt{ripgrep} retrieval tool, yielding a pool of candidate code snippets through exact string matching.

\stitle{Context Construction:} Inspired by GraphCoder~\cite{liu2024graphcoder}, we rank candidate snippets based on their Jaccard similarity~\cite{jaccard1901distribution} with $\mathcal{C}_{local}$. We then select the top-$K$ fragments and concatenate them to form the final prompt context.

\subsection{Baselines}
We compare \modelname against five representative baselines, covering the spectrum from no retrieval to advanced graph-based and learning-based retrieval methods:

\begin{itemize}[leftmargin=*]
    \item \stitle{NoRAG:} The backbone LLM generates code using exclusively intra-file context, without leveraging any external retrieval mechanisms.
    
    \item \stitle{VanillaRAG:} A standard rag baseline employs the BM25 algorithm to assess the lexical similarity between the code under completion and code blocks within the repository. It retrieves the Top-K most relevant code snippets as context to prompt the LLM for the next statement prediction.

    \item \stitle{GraphCoder~\cite{liu2024graphcoder}:} A structure-aware retrieval methodology. This approach constructs a Code Context Graph (CCG) to capture structural dependencies and adopts a coarse-to-fine retrieval strategy, integrating both lexical and structural information to pinpoint relevant code context.

    \item \stitle{RepoFuse~\cite{liang2024repofuse}:} A method designed to address the context-latency conundrum. It integrates analogy context (retrieved via code snippet similarity) and rationale context (derived from import dependency analysis), employing a Rank Truncated Generation (RTG) strategy to select the most relevant cross-file contexts within a constrained window.

    \item \stitle{RLCoder~\cite{wang2024rlcoder}:} An annotation-free reinforcement learning retrieval framework. This approach utilizes the weighted perplexity of code generation as the reward signal to train the retriever.
\end{itemize}

\subsection{Implementation Details}
We evaluate all methods using {DeepSeek-V3.2-EXP} and {Qwen3-Coder-Plus} with the sampling temperature set to 0 for reproducibility. For RAG-based approaches, we standardize the retrieval budget to the Top-$K=10$ code snippets and limit the total context length to 4,096 tokens to ensure a fair comparison. For our method, the number of \texttt{ripgrep} queries is set to $m=10$ via the prompt\add{, while the window size is determined by the LLM}. Methods that require GPU acceleration are executed on a single NVIDIA A6000 GPU, while retrieval latency is measured on the same CPU environment.
\section{Evaluating the Potential of Naive \modelname}
In this chapter, we conduct a systematic empirical study to explore the potential of Naive \modelname in repository-level scenarios. We first evaluate its end-to-end completion performance against sophisticated baselines (RQ1), then dissect the mechanisms behind its success (RQ2), and finally diagnose its limitations to inform directions for optimization (RQ3).

\subsection{RQ1: Evaluating the Performance of Naive \modelname}

\subsubsection{Motivation}
The primary objective of this research question is to quantitatively validate the feasibility of the Naive \modelname framework and to investigate whether simply using LLM-generated grep commands to retrieve code context can effectively improve end-to-end code completion performance. We conduct this preliminary validation on the CrossCodeEval\cite{ding2023crosscodeeval} benchmark.

\subsubsection{Analysis of End-to-End Performance}
Table \ref{tab:performance_comparison} summarizes the performance comparison between Naive \modelname\ and five baselines across the Python and Java subsets. To validate the generalizability of our approach and mitigate potential bias from specific model capabilities, we employ DeepSeek-V3.2-EXP and Qwen3-Coder-Plus as backbone models. Based on the experimental results, we derive the following key findings:

Experimental results show that Naive \modelname  consistently outperforms comparative baselines across all evaluated metrics. In the Python subset under the DeepSeek-V3.2-EXP setting, our method achieves an EM rate of {38.61\%}, substantially exceeding the traditional Vanilla RAG (24.99\%) and RLCoder (36.59\%). Similarly, in the Java subset, Naive \modelname establishes a new benchmark with an EM of 41.70\%. Furthermore, it attains the highest scores on both identifier EM and F1 metrics, indicating that precise character-level matching via \texttt{ripgrep} provides superior accuracy over vector-based retrieval in locating definitions of identifiers, such as functions and variables.

The retrieval time column highlights the practical efficiency of \texttt{ripgrep}. Our method achieves an average retrieval latency of less than 0.02s, outperforming approaches such as GraphCoder and RepoFuse, and substantially reducing the computational cost of the retrieval process.

\begin{table*}[htbp]
    \centering
    \caption{Performance comparison between Naive \modelname\ and baselines on CrossCodeEval}
    \label{tab:performance_comparison}
    \begin{threeparttable}
    \resizebox{0.98\textwidth}{!}{%
    \setlength{\tabcolsep}{2pt}
    \begin{tabular}{@{}llc|cccccccc|cccccccc@{}}
    \toprule
     & & & \multicolumn{8}{c|}{\textbf{DeepSeek-V3.2-EXP}} & \multicolumn{8}{c}{\textbf{Qwen3-Coder-Plus}} \\
    \cmidrule(lr){4-11} \cmidrule(lr){12-19}
     & & \textbf{}
       & \multicolumn{4}{c}{Code} & \multicolumn{4}{c|}{Identifier}
       & \multicolumn{4}{c}{Code} & \multicolumn{4}{c}{Identifier} \\
    \cmidrule(lr){4-7} \cmidrule(lr){8-11} \cmidrule(lr){12-15} \cmidrule(lr){16-19}
     \textbf{Lang} & \textbf{Method} & Retrieval Time(s)
     & EM & ES & Recall & F1 & EM & ES & Recall & F1
     & EM & ES & Recall & F1 & EM & ES & Recall & F1 \\
    \midrule

    \multirow{6}{*}{\textbf{Python}} 
      & No RAG      
      & -- 
      & 15.57 & 66.58 & 82.38 & 81.73 & 22.74 & 66.80 & 57.03 & 55.78
      & 17.45 & 67.77 & 81.46 & 81.99 & 25.55 & 67.84 & 57.21 & 56.86 \\

      & Vanilla RAG 
      & 0.1482 
      & 24.99 & 71.10 & 85.03 & 84.15 & 33.47 & 71.62 & 64.42 & 62.76
      & 27.24 & 72.58 & 84.81 & 84.60 & 36.40 & 73.29 & 64.87 & 64.12 \\

      & GraphCoder  
      & 0.2582 
      & 19.44 & 68.83 & 83.46 & 82.94 & 27.54 & 69.26 & 60.31 & 59.05
      & 21.76 & 69.81 & 83.21 & 83.14 & 30.17 & 70.06 & 60.78 & 60.17 \\

      & RepoFuse    
      & 1.6400 
      & 27.50 & 72.27 & 85.74 & 84.77 & 36.25 & 73.11 & 66.58 & 64.92
      & 29.87 & 73.75 & 85.52 & 85.29 & 39.25 & 74.78 & 67.01 & 66.17 \\

      & RLCoder     
      & -- 
      & {36.59} & {76.92} & {88.90} & {87.27} & {47.32} & {78.01} & {74.23} & {72.16}
      & {40.04} & {79.30} & {88.67} & {88.32} & {51.14} & {80.64} & {75.64} & {74.74} \\

      & Naive \modelname\     
      & 0.0186 
      & \textbf{38.61} & \textbf{77.54} & \textbf{89.08} & \textbf{87.50} & \textbf{48.33} & \textbf{78.67} & \textbf{74.59} & \textbf{72.33}
& \textbf{40.79} & \textbf{79.82} & \textbf{88.96} & \textbf{88.52} & \textbf{51.52} & \textbf{80.97} & \textbf{75.96} & \textbf{74.85} \\

    \midrule

    \multirow{6}{*}{\textbf{Java}} 
      & No RAG      
      & -- 
      & 22.49 & 71.96 & 85.89 & 85.85 & 30.95 & 72.13 & 65.13 & 64.11
      & 21.79 & 70.87 & 83.79 & 84.30 & 30.39 & 70.82 & 62.74 & 62.55 \\

      & Vanilla RAG 
      & 0.1015 
      & 29.64 & 74.96 & 87.50 & 87.44 & 39.46 & 75.51 & 69.65 & 68.67
      & 28.99 & 73.64 & 85.51 & 85.94 & 37.54 & 73.77 & 67.28 & 66.91 \\

      & GraphCoder  
      & 0.1110 
      & 25.20 & 73.09 & 86.19 & 86.26 & 34.27 & 73.26 & 66.77 & 65.95
      & 24.12 & 71.44 & 84.31 & 84.75 & 32.40 & 71.48 & 63.83 & 63.72 \\

      & RepoFuse    
      &  0.0938
      & 38.62 & 78.36 & 89.23 & 89.12 & 50.35 & 79.34 & 75.54 & 74.65
      & 39.97 & 77.62 & 87.43 & 87.91 & 50.44 & 78.36 & 73.72 & 73.57 \\

      & RLCoder     
      & -- 
      & {39.46} & {78.84} & {89.60} & {89.22} & {51.24} & {79.87} & {76.55} & {75.24}
      & {41.19} & {78.13} & {87.70} & {88.20} & {51.38} & {78.89} & {74.58} & {74.35} \\

      & Naive \modelname\     
      & 0.0173
      & \textbf{41.70} & \textbf{78.93} & \textbf{89.93} & \textbf{89.33} & \textbf{52.17} & \textbf{79.95} & \textbf{76.62} & \textbf{75.67}
& \textbf{41.42} & \textbf{78.31} & \textbf{87.92} & \textbf{88.67} & \textbf{51.90} & \textbf{79.02} & \textbf{74.69} & \textbf{74.77} \\

    \bottomrule
    \end{tabular}%
    }
    \begin{tablenotes}
            \tiny
            \item[*] Retrieval time for RLCoder is not reported due to its reliance on GPU-based inference.
        \end{tablenotes}
    \end{threeparttable}
\end{table*}

\subsubsection{Analysis of Solved Instance Overlap}
\begin{figure}[tbp]
  \centering
  
  \begin{subfigure}{0.42\textwidth}
    \centering
    \includegraphics[width=\linewidth]{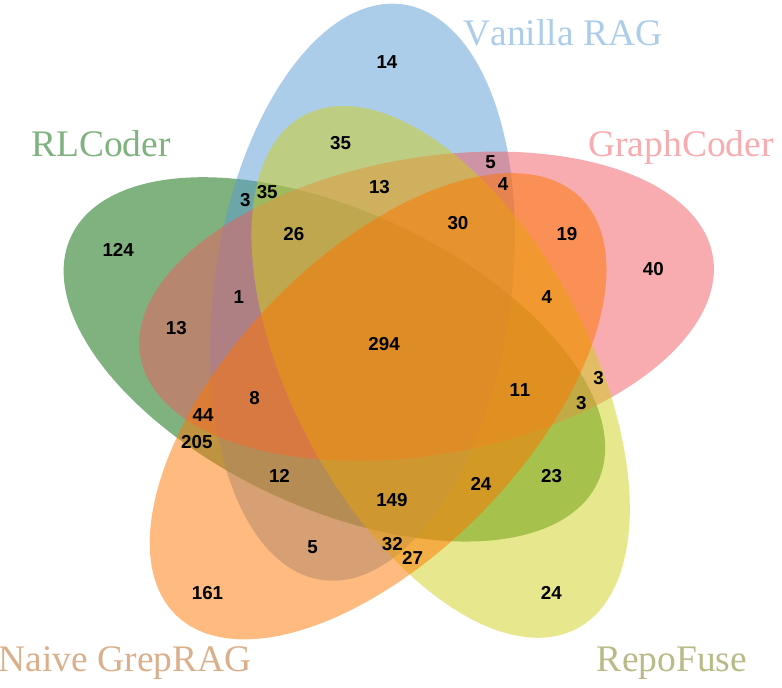}
    \caption{Python subset of CrossCodeEval}
    \label{fig:venn-python}
  \end{subfigure}
  \hspace{1em}
  \begin{subfigure}{0.42\textwidth}
    \centering
    \includegraphics[width=\linewidth]{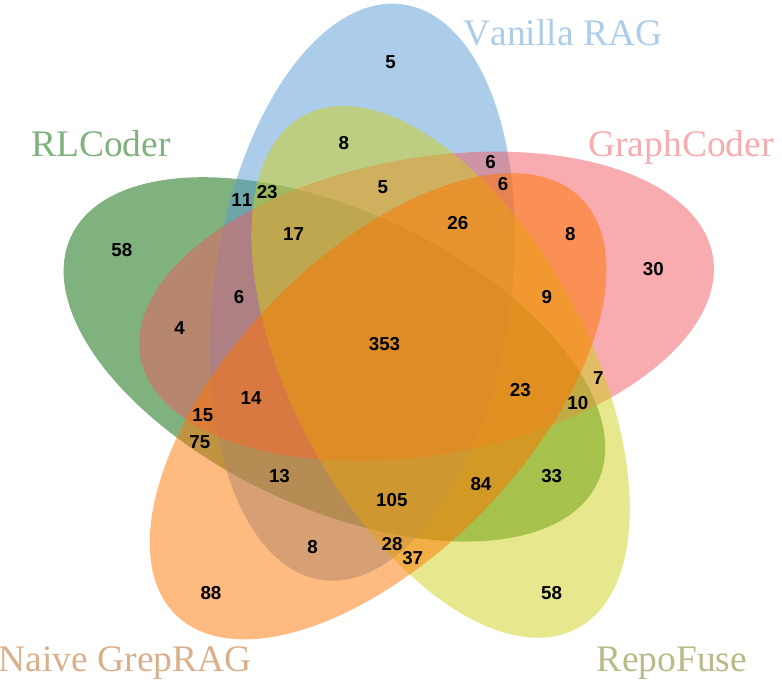}
    \caption{Java subset of CrossCodeEval}
    \label{fig:venn-java}
  \end{subfigure}

  \caption{
    Comparison of distinct solved cases across systems in the CrossCodeEval dataset.
  }
  \label{fig:venn-comparison}
  \vspace{-0.4cm}
\end{figure}

We utilize a Venn diagram (Figure~\ref{fig:venn-comparison}) to visually compare the solution sets of Naive \modelname, VanillaRAG, GraphCoder~\cite{liu2024graphcoder}, RepoFuse~\cite{liang2024repofuse}, and RLCoder~\cite{wang2024rlcoder} under EM (EM=100\%), using DeepSeek-V3.2-EXP as the underlying LLM. In the Python dataset, \modelname uniquely resolved \textbf{161} instances where baselines failed, a figure significantly higher than the number of instances unique to RepoFuse (24). The Java dataset exhibits a similar trend, confirming that explicit literal retrieval effectively captures precise keyword matches often overlooked by semantic models.
Furthermore, we observe that instances successfully solved by all models simultaneously account for only 20\%--30\% of the total (Python: 294/1391, Java: 353/1173), indicating a substantial divergence in solution spaces. Notably, \modelname contributed the largest unique solution set (accounting for 11.6\% in Python). This phenomenon not only validates the complementarity between different retrieval strategies but also suggests that a subset of code completion tasks inherently relies on explicit lexical pattern matching.

\subsection{RQ2: Analyzing the Success of Naive \modelname}

\subsubsection{Motivation} 
While RQ1 has shown the effectiveness of Naive \modelname, the reasons behind its success remain unclear.
We analyze its internal retrieval patterns and examine why it succeeds where other RAG-based baselines fail.

\subsubsection{Analysis of Retrieval Patterns}
To characterize retrieval patterns, we analyze $N=45,615$ \texttt{ripgrep} commands (25,366 Python, 20,249 Java). Using a 95\% confidence level and a 5\% margin of error, we perform stratified random sampling by programming language, yielding a sample of 381 commands (212 Python, 169 Java). 
Our analysis reveals that Naive \modelname\ retrieval behaviors can be understood at two levels: basic retrieval patterns and advanced retrieval strategies.

\stitle{Basic Retrieval Patterns}
At the individual command level, the keywords are primarily categorized into four types, each keyword type corresponds to a specific information retrieval target and plays a role in different code completion scenarios.

\begin{itemize}[leftmargin=*]

\item \stitle{Class-name Retrieval (35.96\%)}
When a \texttt{ripgrep} command uses a class name, its main goal is to reveal the object’s type and member structure, which is crucial in the following two completion scenarios:
(1) Method call completion (\texttt{obj.}): The LLM uses the object’s class name to retrieve its definition, directly accessing its attributes and methods to guide member completion.
(2) Class declaration completion (\texttt{class C extends P}): When a class inherits from a parent, the LLM uses the parent class name to retrieve its definition, obtaining attributes and methods as structural references for completing the subclass.

\item  \stitle{Method-name Retrieval (41.47\%)} 
\texttt{ripgrep} commands that use method names as keywords are primarily used to locate the definition and usage of the target method within the codebase, providing direct references for completing calls or implementations. 
(1) Arguments completion (\texttt{obj.method(...)}): The LLM uses the method identifier to retrieve its signature across the repository, obtaining parameters and return type, and also retrieves call examples elsewhere in the project to guide argument filling and usage patterns.
(2) Method body completion (\texttt{def method(...):}): When a method is only provided with a function signature but lacks a concrete implementation, the LLM retrieves same-name or fuzzy-matched methods to reference their internal logic, assisting in implementing the target method.

\item  \stitle{Variable-name Retrieval (18.37\%)} 
This retrieves variable definitions and assignments to understand types, initial values, and usage. Global variables such as \texttt{CONFIG\_PATH} are traced to their definitions for configuration or state information, while frequent local variables appearing near the code region being completed are retrieved to provide usage references for completion.

\item  \stitle{Others (4.20\%)} 
A small portion uses strings or non-standard identifiers to locate similar code fragments, which the model references to aid target code completion.

\end{itemize}

\stitle{Advanced Retrieval Strategies}
Beyond keyword matching, Naive \modelname\ exhibits sophisticated behaviors by enhancing single-query recall and orchestrating multiple queries for a holistic view.
\begin{itemize}[leftmargin=*]
    \item \stitle{Fuzzy Matching via Wildcards.} 
    Notably, 23.5\% of the sampled commands utilize wildcard patterns. Rather than strict identity matching, these queries aim to retrieve code snippets with similar naming patterns (e.g., \texttt{class.*ConfigModel} matching implementations like \texttt{DataConfigModel}). This allows the model to reference these semantically related definitions as prototypes, facilitating completion by mimicking their structure even when the exact identifier in the target context is unknown or lacks a direct counterpart.
    
    \item \stitle{Multi-Query Retrieval.} 
    A key characteristic of Naive \modelname\ is its ability to generate a {query set} rather than relying on isolated commands. For example, when performing argument completion (\texttt{obj.method(...)}), class-name queries retrieve the receiver’s class definition, exposing available method signatures and implementations. Method-name queries help locate the target method and its callsites, revealing method parameters and return types. Variable-name queries surface other usages of the same variable, offering additional clues about expected arguments and return values. By aggregating these results, Naive \modelname\ leverages multiple, partially overlapping lexical views to facilitate code completion, significantly improving retrieval robustness and reducing reliance on any single query.
\end{itemize}

\subsubsection{Failure Analysis of RAG-based Baselines}
The above retrieval patterns primarily rely on lexical matching of explicit identifiers. While conceptually simple, they can successfully solve a subset of test cases that other RAG-based baselines fail to solve. We analyze the set $S_{unique}$, which consists of 249 test cases (161 Python, 88 Java) that are correctly completed by Naive \modelname\ (EM = 100\%), in order to investigate why these baselines fail on these cases.

Despite differences in their implementation, these baselines share a two-stage workflow: initial lexical retrieval (BM25/Jaccard), followed by re-ranking mechanism. Specifically, GraphCoder and RLCoder first perform coarse retrieval using Jaccard or BM25, then refine with structural or semantic similarity; RepoFuse merges BM25 Top-$k$ results with structurally retrieved code blocks before final re-ranking. Based on this workflow, we categorize the failures of baselines on $S_{unique}$ into two mutually exclusive types.

\begin{itemize}[leftmargin=*]
\item \textbf{Type I: Coarse Retrieval Failure.}
In this type, baselines fail to include the critical context in the initial lexical retrieval stage, indicating that the failure occurs at the front end of the retrieval pipeline, before any baseline-specific re-ranking takes place.

\item \textbf{Type II: Re-ranking Failure.}
In this type, baselines retrieve core code blocks in coarse retrieval, \delete{but re-ranking fails to prioritize them.}\add{but still fail to generate the correct completion, indicating deficiencies in the re-ranking stage.}
\end{itemize}

Given that the context retrieved by RAG methods typically comprises multiple discrete code blocks, we quantify the coverage of critical context by baselines' retrieval results using a threshold-based set overlap metric. Here, the set of code fragments retrieved by Naive \modelname\ for each sample serves as the {golden context}, denoted $\mathcal{C}_{gold}$. For each baseline, we define the coverage of $\mathcal{C}_{gold}$ by its retrieved set $\mathcal{C}_{retrieved}$ as the line-level intersection ratio:
\begin{equation}
    I(\mathcal{C}_{retrieved}, \mathcal{C}_{gold}) = \frac{|Lines(\mathcal{C}_{retrieved}) \cap Lines(\mathcal{C}_{gold})|}{|Lines(\mathcal{C}_{gold})|}
\end{equation}

We set the determination threshold at $\tau=0.8$, meaning a retrieval result effectively recalls the core information if it covers more than 80\% of the code lines in the golden context. This threshold accommodates minor boundary discrepancies introduced by different chunking granularities while ensuring that the essential informational content is preserved. A sensitivity analysis over various thresholds confirmed consistent trends, for simplicity, we only report results for $\tau=0.8$. \add{Among the failed completion cases, we classify a sample as a coarse retrieval failure when 
\(I(\mathcal{C}_{retrieved}, \mathcal{C}_{gold}) < \tau\),
and as a re-ranking failure when 
\(I(\mathcal{C}_{retrieved}, \mathcal{C}_{gold}) \ge \tau\).}

As shown in Table~\ref{tab:rq2_failure_analysis_percent}, most baseline failures on $S_{unique}$ are due to coarse retrieval. This failure is mainly caused by the limitations of global lexical similarity metrics, such as BM25 and Jaccard, which emphasize overall token overlap rather than identifiers closely related to the completion site. This indicates that current RAG methods relying on BM25 or Jaccard for coarse retrieval can be problematic for code completion. In contrast, Naive \modelname\ retrieves explicit identifiers at the completion site, allowing precise recall of locally and structurally relevant code. For example, in member-access completion (\texttt{obj.}), it reliably retrieves the class definition of \texttt{obj}. Under BM25- or Jaccard-based coarse retrieval, such class definitions often exhibit weak global lexical overlap with the surrounding context and are therefore filtered out before any re-ranking can take place. For RepoFuse, although it incorporates structural dependencies, its coarse retrieval stage still lacks explicit matching of local variable or method identifiers, leading to similar recall failures.

We next analyze re-ranking failures, which primarily stem from the baselines' inability to prioritize precise identifier matches. 
Specifically, GraphCoder emphasizes structural similarity (e.g., similar loop patterns), which can be irrelevant to the actual completion target. RLCoder uses a fine-tuned retriever to encode semantic similarity. While it captures conceptually related code, it often fails to rank exact method or variable names highly and its results are less interpretable. RepoFuse combines BM25-based lexical retrieval with call-chain–based dependency signals. However, its re-ranking stage still lacks the precision of Naive \modelname\ in emphasizing local identifiers, resulting in weak attention to identifiers with identical names.

Overall, Naive \modelname succeeds because it retrieves code fragments more closely related to the completion site and uses more precise query keywords, whereas baselines fail due to inaccurate initial retrieval and re-ranking failures.

\begin{table}[t]
\centering
\small 
\setlength{\tabcolsep}{4pt} 
\caption{Breakdown of failure modes on $S_{unique}$.}
\label{tab:rq2_failure_analysis_percent}
\begin{tabular}{l c c c c}
\toprule
\multirow{2}{*}[-1ex]{\textbf{Method}} & \multicolumn{2}{c}{\textbf{Python (\%)}} & \multicolumn{2}{c}{\textbf{Java (\%)}} \\
\cmidrule(lr){2-3} \cmidrule(lr){4-5}
 & Recall Failure & Re-ranking Failure & Recall Failure & Re-ranking Failure \\
\midrule
GraphCoder & 73.3 & 26.7 & 76.1 & 23.9 \\
RLCoder    & 68.9 & 31.1 & 70.5 & 29.5 \\
RepoFuse   & 64.6 & 35.4 & 65.9 & 34.1 \\
\bottomrule
\end{tabular}
\vspace{-0.4cm}
\end{table}
\subsection{RQ3: Limitations of Naive \modelname}

\subsubsection{Motivation}
In the Venn analysis of RQ1, we observe that the solution space of Naive \modelname\ does not fully cover that of the baselines, which means there exist scenarios where a baseline succeeds while Naive \modelname\ fails. This research question aims to investigate the limitations of Naive \modelname\ and provide insights for subsequent improvements.

\subsubsection{Methodology}
We construct a failure dataset, denoted as $S_{fail}$, comprising test cases where Naive \modelname fails to generate a correct prediction (EM=0\%) whereas at least one of the other RAG-based baselines succeeded. We conduct a quantitative census of all samples meeting these criteria (362 Python and 281 Java cases).
We focus on the actual retrieval pipeline of Naive \modelname. Following the classification approach in RQ2, we categorize failures into recall failure and re-ranking failure. \add{Here, the golden context denotes the retrieved code fragments from the baseline that successfully completes the sample, and their ranking is used as the reference.} Recall failure occurs when the Naive \modelname\ fails to retrieve the golden context by the ripgrep command set, whereas re-ranking failure occurs when the golden context is successfully retrieved but \delete{not assigned a sufficiently high rank.}\add{the completion is still incorrect.}

\subsubsection{Analysis Results}
Based on the aforementioned classification criteria, we analyze 643 failure samples. The data reveals that re-ranking failure is the dominant factor, accounting for approximately 71.5\% of Python cases and 75.1\% of Java cases, while recall failure constitutes the remaining 28.5\% and 24.9\%, respectively. We perform open coding on these samples to identify typical failure scenarios. To ensure coding rigor, the first two authors independently annotated the samples, achieving a Cohen's kappa of 0.82, indicating substantial agreement. Through this process, we categorize failures into three representative classes: two reflecting distinct patterns of re-ranking failure, and one corresponding to the fundamental limitation of recall failure.

\stitle{Keyword Ambiguity and Re-ranking Failure}: 
    When queries involve high-frequency generic identifiers such as \texttt{init}, \texttt{config}, or \texttt{run}, \texttt{ripgrep} retrieves a large number of irrelevant documents containing identical keywords but lacking semantic relevance. In the presence of such noise, current Naive \modelname relies on Jaccard similarity to compute the token overlap rate between retrieved chunks and the query. This approach cannot effectively distinguish frequent stop words from task-specific identifiers. Consequently, noise chunks containing numerous frequent terms accrue artificially high Jaccard scores, displacing the code blocks that actually contain the relevant context from the Top-K candidates.

\begin{figure*}[htbp] \centering \includegraphics[width=\textwidth]{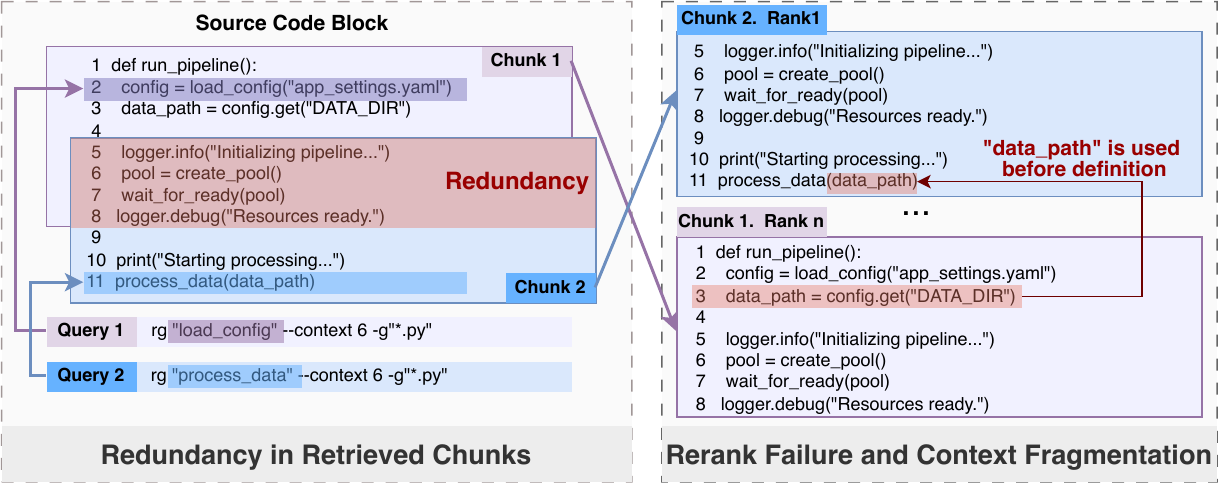} 
\caption{Redundancy and Context Fragmentation. Two independent grep queries hit adjacent regions within the same file.} 
\label{fig:fragmentation_example} 
\vspace{-0.4cm}
\end{figure*}

\stitle{Context Fragmentation and Redundancy}: 
    Even when the correct context successfully enters the Top-K, its structural organization often exhibits issues. As illustrated in Figure~\ref{fig:fragmentation_example}, two independent ripgrep queries targeting distinct keywords (\texttt{"load\_config"} and \texttt{"process\_data"}) respectively matched adjacent regions within the same file. The independent retrieval mechanism of grep primarily induces {information redundancy}, resulting in the duplicate retention of the overlapping code region (L5--8) in the final input, wastefully consuming the finite token budget. Furthermore, this mechanism precipitates {semantic discontinuity}, where the snippet containing the variable \texttt{data\_path}'s {usage site} (Chunk 2) obtains a higher relevance score than its {definition site} (Chunk 1). This results in a sequence where usage precedes definition in the context ultimately fed to the LLM. Such physical fragmentation and chronological disorder disrupt the logical flow of the code, exacerbating the difficulty for the LLM to comprehend the context.

\stitle{Implicit Dependencies}: 
    It represents a fundamental limitation of lexical retrieval. The failures mainly arise when the current context lacks explicit structural relationships, such as inheritance, making it difficult for the Naive \modelname to locate the critical context using ripgrep commands.

\section{Improving Naive \modelname\ via Post-Processing}

\subsection{Overview}
RQ3 reveals that although Naive \modelname exhibits an exceptionally high recall rate, the primitive Jaccard re-ranking mechanism is inadequate for addressing keyword ambiguity, resulting in distractor documents with identical keywords but low semantic relevance. Furthermore, the absence of a deduplication mechanism leads to context redundancy and fragmentation. Consequently, this section investigates whether an efficient post-processing module can be introduced to resolve ranking bottlenecks and redundancy issues.

\subsection{Approach}

\begin{figure*}[tbp] 
\centering \includegraphics[width=\textwidth]{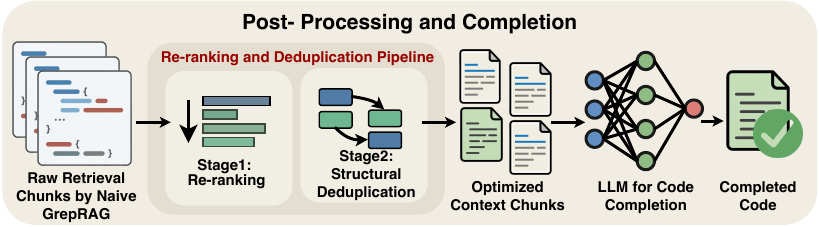} 
\caption{Post-processing pipeline built upon the Naive \modelname\ framework.}
\label{fig:model} 
\end{figure*}

As shown in Figure~\ref{fig:model}, our approach builds upon the Naive \modelname pipeline by retaining the original \texttt{ripgrep}-based retrieval process and introducing two cascaded post-processing steps applied to the retrieved chunks.

\stitle{Identifier-Weighted Re-ranking}. 
    To address the keyword ambiguity issue analyzed previously, we require an algorithm that effectively penalizes frequent generic identifiers while rewarding low-frequency, task-specific identifiers. BM25 introduces an IDF factor that suppresses the contribution of frequent terms and relatively amplifies the weight of rare identifiers. Consequently, replacing the re-ranking strategy with BM25 yields a more principled and discriminative ranking. For each chunk $C_i$ retrieved by Grep, we regard it as a document and the code under completion as the query to calculate a relevance score. This step outputs a candidate list $L_{ranked}$, sorted in descending order of relevance. This does not contradict our findings in RQ2. This highlights that BM25 is effective for assigning differential weights in a completion-aware candidate set, but may be unsuitable as a coarse, global retriever in code completion.

\stitle{Structure-Aware De-duplication and Fusion}. 
    To mitigate token wastage and semantic discontinuity, we devise a fusion strategy based on line number intervals. This mechanism parses the physical line number range of each chunk to precisely identify physically overlapping or adjacent code snippets. Subsequently, we execute a concatenation operation to merge fragmented snippets into complete, contiguous semantic blocks, thereby reconstructing the logical flow of the code while eliminating redundancy. To balance computational overhead with performance, we process only the Top-N\% (set to 50\% in our experiments) of candidate blocks from $L_{ranked}$.
    
    Finally, we select the Top-K blocks from the de-duplicated list to serve as input to the LLM, while strictly limiting the total context length to 4,096 tokens.

\subsection{Main Results}

Table \ref{tab:merged-performance} presents the primary performance metrics of \modelname\ across two datasets. The results indicate that \modelname achieves substantial improvements across almost all evaluation dimensions on the CrossCodeEval~\cite{ding2023crosscodeeval} dataset. Furthermore, these gains maintain high consistency across different backbone models. For Python completion tasks using DeepSeek-V3.2-EXP as the backbone, \modelname\ improves the code EM from 38.61\% (Naive version) to 42.29\%, and increases the identifier F1 from 72.33 to 75.15. These results significantly outperform baseline models such as RepoFuse, establishing a new SOTA. Notably, this performance enhancement is not confined to specific backbone models; when employing Qwen3-Coder-Plus for completion, \modelname\ similarly propels the code EM on Python tasks to 44.62\%, significantly surpassing both the baselines and Naive \modelname. Overall, across different tasks and backbone models on CrossCodeEval, \modelname\ improves code EM by 7.04\%–15.58\% and identifier EM by 5.02\%–11.50\% relative to the best-performing baselines. This confirms that our framework does not rely on the parameter preferences of specific models, but rather provides a generalized context augmentation capability.

The RepoEval\_Updated~\cite{liu2024graphcoder} dataset consists of repositories with a significantly larger code volume. In this scenario, \modelname\ demonstrates exceptional noise robustness. Particularly in API-level tasks, the model performance improves significantly compared to the Naive version. On the Python subset, code EM increases by 13.8\% (35.75 $\rightarrow$ 40.70), and on the Java subset by 13.4\% (40.27 $\rightarrow$ 45.67). This trend is similarly observed on the Qwen3-Coder-Plus model.

\begin{table*}[htbp]
    \centering
    \caption{Performance comparison on CrossCodeEval and RepoEval\_Updated}
    \label{tab:merged-performance}
    \begin{threeparttable}
    \resizebox{0.98\textwidth}{!}{%
    \setlength{\tabcolsep}{2pt}
    \begin{tabular}{@{}lllc c|cccccccc|cccccccc@{}}
    \toprule
    & & & {Retrieval} & {RAG Pipeline} & \multicolumn{8}{c|}{\textbf{DeepSeek-V3.2-EXP}} & \multicolumn{8}{c}{\textbf{Qwen3-Coder-Plus}} \\
    \cmidrule(lr){6-13} \cmidrule(lr){14-21}
    & & & {Time(s)} & {Time(s)} & \multicolumn{4}{c}{Code} & \multicolumn{4}{c|}{Identifier} & \multicolumn{4}{c}{Code} & \multicolumn{4}{c}{Identifier} \\
    \cmidrule(lr){6-9} \cmidrule(lr){10-13} \cmidrule(lr){14-17} \cmidrule(lr){18-21}
    \textbf{Dataset} & \textbf{Task-Lang} & \textbf{Method} & (s) & (s) & EM & ES & Recall & F1 & EM & ES & Recall & F1 & EM & ES & Recall & F1 & EM & ES & Recall & F1 \\
    \midrule

    \multirow{16}{*}{\rotatebox{90}{\textbf{CrossCodeEval}}}
    & \multirow{8}{*}{\makecell{Line-\\Python}} 
      & No RAG & -- & -- & 15.57 & 66.58 & 82.38 & 81.73 & 22.74 & 66.80 & 57.03 & 55.78 & 17.45 & 67.77 & 81.46 & 81.99 & 25.55 & 67.84 & 57.21 & 56.86 \\
    & & Vanilla RAG & 0.1482 & 0.36 & 24.99 & 71.10 & 85.03 & 84.15 & 33.47 & 71.62 & 64.42 & 62.76 & 27.24 & 72.58 & 84.81 & 84.60 & 36.40 & 73.29 & 64.87 & 64.12 \\
    & & GraphCoder & 0.2582 & 1.56 & 19.44 & 68.83 & 83.46 & 82.94 & 27.54 & 69.26 & 60.31 & 59.05 & 21.76 & 69.81 & 83.21 & 83.14 & 30.17 & 70.06 & 60.78 & 60.17 \\
    & & RepoFuse & 1.6400 & 3.29 & 27.50 & 72.27 & 85.74 & 84.77 & 36.25 & 73.11 & 66.58 & 64.92 & 29.87 & 73.75 & 85.52 & 85.29 & 39.25 & 74.78 & 67.01 & 66.17 \\
    & & RLCoder & -- & -- & 36.59 & 76.92 & 88.90 & 87.27 & 47.32 & 78.01 & 74.23 & 72.16 & 40.04 & 79.30 & 88.67 & 88.32 & 51.14 & 80.64 & 75.64 & 74.74 \\
    & & Naive \modelname & 0.0186 & 5.04 & 38.61 & 77.54 & 89.08 & 87.50 & 48.33 & 78.67 & 74.59 & 72.33 & 40.79 & 79.82 & 88.96 & 88.52 & 51.52 & 80.97 & 75.96 & 74.85 \\
    & & \modelname & 0.0197 & 5.04 & \textbf{42.29} & \textbf{79.66} & \textbf{89.53} & \textbf{88.50} & \textbf{52.76} & \textbf{80.76} & \textbf{76.84} & \textbf{75.15} & \textbf{44.62} & \textbf{81.32} & \textbf{89.58} & \textbf{89.45} & \textbf{55.87} & \textbf{82.62} & \textbf{77.90} & \textbf{77.29} \\
    & & {\modelname (0.6b)} 
& {0.0191} 
& {1.31} 
& {{40.15}} 
& {{79.15}} 
& {{89.52}} 
& {{88.06}} 
& {{51.82}} 
& {{80.03}} 
& {{76.01}} 
& {{74.44}} 
& {{41.01}} 
& {{79.89}} 
& {{88.98}} 
& {{88.63}} 
& {{51.63}} 
& {{81.09}} 
& {{76.02}} 
& {{74.98}} \\
    \cmidrule(lr){2-21}

    & \multirow{8}{*}{\makecell{Line-\\Java}} 
      & No RAG & -- & -- & 22.49 & 71.96 & 85.89 & 85.85 & 30.95 & 72.13 & 65.13 & 64.11 & 21.79 & 70.87 & 83.79 & 84.30 & 30.39 & 70.82 & 62.74 & 62.55 \\
    & & Vanilla RAG & 0.1015 & 0.41 & 29.64 & 74.96 & 87.50 & 87.44 & 39.46 & 75.51 & 69.65 & 68.67 & 28.99 & 73.64 & 85.51 & 85.94 & 37.54 & 73.77 & 67.28 & 66.91 \\
    & & GraphCoder & 0.1110 & 1.73 & 25.20 & 73.09 & 86.19 & 86.26 & 34.27 & 73.26 & 66.77 & 65.95 & 24.12 & 71.44 & 84.31 & 84.75 & 32.40 & 71.48 & 63.83 & 63.72 \\
    & & RepoFuse & 0.0938 & 1.53 & 38.62 & 78.36 & 89.23 & 89.12 & 50.35 & 79.34 & 75.54 & 74.65 & 39.97 & 77.62 & 87.43 & 87.91 & 50.44 & 78.36 & 73.72 & 73.57 \\
    & & RLCoder & -- & -- & 39.46 & 78.84 & 89.60 & 89.22 & 51.24 & 79.87 & 76.55 & 75.24 & 41.19 & 78.13 & 87.70 & 88.20 & 51.38 & 78.89 & 74.58 & 74.35 \\
    & & Naive \modelname & 0.0173 & 4.54 & 41.70 & 78.93 & \textbf{89.93} & 89.33 & 52.17 & 79.95 & 76.62 & 75.67 & 41.42 & 78.31 & 87.92 & 88.67 & 51.90 & 79.02 & 74.69 & 74.77 \\
    & & \modelname & 0.0181 & 4.54 & \textbf{43.15} & \textbf{80.07} & {89.78} & \textbf{89.83} & \textbf{53.81} & \textbf{80.88} & \textbf{77.41} & \textbf{76.57} & \textbf{44.09} & \textbf{79.95} & \textbf{88.98} & \textbf{89.29} & \textbf{54.09} & \textbf{80.69} & \textbf{76.83} & \textbf{76.57} \\
    & & {\modelname (0.6b)} 
& {0.0176} 
& {1.42} 
& {{42.87}} 
& {{79.96}} 
& {89.62} 
& {{89.78}} 
& {{53.76}} 
& {{80.74}} 
& {{77.18}} 
& {{76.33}} 
& {{41.61}} 
& {{78.44}} 
& {{88.12}} 
& {{88.69}} 
& {{52.17}} 
& {{79.38}} 
& {{74.77}} 
& {{74.95}} \\
    \midrule

    \multirow{32}{*}{\rotatebox{90}{\textbf{RepoEval\_Updated}}}
    & \multirow{8}{*}{\makecell{Line-\\Python}} 
      & No RAG & -- & -- & 34.25 & 64.29 & 82.79 & 80.36 & 40.80 & 66.16 & 59.09 & 56.15 & 51.20 & 74.02 & 85.58 & 85.14 & 56.50 & 75.22 & 67.82 & 66.64 \\
    & & Vanilla RAG & 10.04 & 12.14 & 36.90 & 65.20 & 83.33 & 80.71 & 43.25 & 66.85 & 60.26 & 57.56 & 52.40 & 75.12 & 86.81 & 85.95 & 57.80 & 76.14 & 69.49 & 68.12 \\
    & & GraphCoder & 7.63 & >60 & 43.70 & 67.95 & 83.30 & 81.42 & 49.50 & 69.94 & 63.28 & 61.05 & 56.70 & 77.41 & 87.92 & 87.32 & 62.10 & 78.68 & 71.98 & 70.81 \\
    & & RepoFuse & 23.03 & >60 & 36.85 & 65.71 & 84.43 & 81.41 & 43.65 & 67.61 & 60.96 & 57.95 & 51.30 & 74.00 & 86.36 & 85.39 & 56.80 & 75.31 & 68.79 & 67.51 \\
    & & RLCoder & -- & -- & 42.60 & 68.82 & 84.95 & 82.48 & 48.85 & 70.61 & 64.96 & 61.89 & 57.70 & 77.91 & 88.24 & 87.63 & 62.70 & 78.76 & 72.02 & 70.88 \\
    & & Naive \modelname & 0.51 & 5.63 & 41.45 & 68.14 & 85.33 & 82.88 & 47.85 & 69.72 & 63.09 & 60.36 & 57.65 & 77.43 & 87.84 & 87.69 & 62.10 & 77.69 & 71.26 & 69.83 \\
    & & \modelname & 0.62 & 5.74 & {44.90} & {69.92} & {86.49} & {83.74} & {50.98} & {71.78} & {65.67} & {62.56} & {59.05} & {78.27} & {88.49} & {87.74} & {64.10} & {79.58} & {72.54} & {71.60} \\
    & & {\modelname (0.6b)} 
& {0.55} 
& {1.93} 
& {\textbf{44.95}} 
& {\textbf{69.98}} 
& {\textbf{86.67}} 
& {\textbf{83.93}} 
& {\textbf{51.40}} 
& {\textbf{71.92}} 
& {\textbf{65.78}} 
& {\textbf{62.90}} 
& {\textbf{59.15}} 
& {\textbf{78.30}} 
& {\textbf{88.53}} 
& {\textbf{87.82}} 
& {\textbf{64.30}} 
& {\textbf{79.66}} 
& {\textbf{72.61}} 
& {\textbf{71.73}} \\
    \cmidrule(lr){2-21}

    & \multirow{8}{*}{\makecell{Line-\\Java}} 
      & No RAG & -- & -- & 34.25 & 64.29 & 82.79 & 80.36 & 40.80 & 66.16 & 59.09 & 56.15 & 45.55 & 75.87 & 86.16 & 86.31 & 55.00 & 77.06 & 70.48 & 69.93 \\
    & & Vanilla RAG & 15.04 & 16.84 & 35.00 & 67.22 & 82.88 & 81.73 & 44.20 & 69.21 & 63.07 & 60.81 & 48.40 & 77.03 & 87.14 & 87.16 & 57.50 & 77.91 & 71.80 & 71.17 \\
    & & GraphCoder & 10.49 & >60 & 39.30 & 69.59 & 83.35 & 82.67 & 48.05 & 71.17 & 64.58 & 62.89 & 51.40 & 78.63 & 87.72 & 87.97 & 60.05 & 79.66 & 72.94 & 72.50 \\
    & & RepoFuse & 1.96 & >60 & 33.95 & 66.87 & 83.08 & 81.54 & 42.95 & 68.02 & 62.57 & 60.10 & 48.70 & 77.03 & 86.62 & 86.92 & 57.40 & 77.86 & 71.59 & 71.12 \\
    & & RLCoder & -- & -- & 40.10 & 70.67 & 85.16 & 83.88 & 49.05 & 72.34 & 67.05 & 64.69 & 53.00 & 79.05 & 88.15 & 88.26 & 61.50 & 80.00 & 74.03 & 73.48 \\
    & & Naive \modelname & 0.26 & 5.30 & 40.50 & 70.66 & 85.01 & 83.88 & 49.95 & 72.94 & 66.66 & 64.57 & 53.67 & 79.01 & 88.03 & 88.28 & 62.05 & \textbf{80.74} & 73.97 & 73.20 \\
    & & \modelname & 0.28 & 5.32 & {43.65} & {72.14} & {85.92} & {84.55} & {52.40} & {73.90} & {68.83} & {66.41} & {54.55} & {79.29} & {88.18} & {88.92} & {63.35} & 80.61 & {74.50} & {73.85} \\
    & & {\modelname (0.6b)} 
& {0.26} 
& {1.69} 
& {\textbf{44.95}} 
& {\textbf{73.23}} 
& {\textbf{86.39}} 
& {\textbf{85.14}} 
& {\textbf{53.35}} 
& {\textbf{75.15}} 
& {\textbf{69.75}} 
& {\textbf{67.35}} 
& {\textbf{54.65}} 
& {\textbf{79.32}} 
& {\textbf{88.26}} 
& {\textbf{88.95}} 
& {\textbf{63.45}} 
& {{80.71}} 
& {\textbf{74.55}} 
& {\textbf{73.91}} \\
    \cmidrule(lr){2-21}

    & \multirow{8}{*}{\makecell{API-\\Python}} 
      & No RAG & -- & -- & 34.40 & 65.80 & 83.13 & 83.22 & 38.10 & 67.22 & 61.96 & 61.28 & 48.35 & 72.95 & 84.53 & 85.99 & 51.10 & 73.54 & 69.14 & 69.87 \\
    & & Vanilla RAG & 11.78 & 14.75 & 32.65 & 63.79 & 81.81 & 81.95 & 37.15 & 65.04 & 59.72 & 58.77 & 47.60 & 72.02 & 84.14 & 85.70 & 50.40 & 72.57 & 67.78 & 68.68 \\
    & & GraphCoder & 7.54 & >60 & 40.60 & 67.33 & 82.63 & 83.04 & 44.60 & 68.55 & 64.26 & 63.78 & 52.65 & 75.98 & 86.30 & 87.89 & 55.55 & 76.51 & 72.63 & 73.61 \\
    & & RepoFuse & 50.99 & >60 & 35.40 & 66.36 & 83.70 & 83.52 & 40.20 & 67.75 & 63.83 & 62.23 & 49.45 & 73.65 & 85.14 & 86.53 & 52.45 & 74.22 & 69.84 & 70.60 \\
    & & RLCoder & -- & -- & 39.85 & 67.64 & 83.45 & 83.60 & 44.05 & 69.12 & 65.39 & 64.05 & 52.65 & 75.40 & 85.78 & 87.43 & 55.45 & 76.12 & 71.46 & 72.44 \\
    & & Naive \modelname & 0.39 & 5.53 & 35.75 & 65.36 & 82.86 & 82.79 & 40.05 & 66.70 & 61.84 & 60.84 & 50.97 & 75.49 & 85.12 & 86.27 & 53.23 & 74.45 & 70.02 & 70.87 \\
    & & \modelname & 0.49 & 5.63 & {40.70} & {68.86} & {84.61} & {84.50} & {45.15} & {70.07} & {65.98} & {64.91} & {53.35} & {76.18} & {86.94} & {88.36} & {56.40} & \textbf{77.07} & {73.18} & {73.83} \\
    & & {\modelname (0.6b)} 
& {0.41} 
& {2.03} 
& {\textbf{44.90}} 
& {\textbf{70.88}} 
& {\textbf{84.97}} 
& {\textbf{85.44}} 
& {\textbf{49.55}} 
& {\textbf{71.93}} 
& {\textbf{67.95}} 
& {\textbf{67.68}} 
& {\textbf{53.50}} 
& {\textbf{76.29}} 
& {\textbf{87.06}} 
& {\textbf{88.42}} 
& {\textbf{56.55}} 
& {{77.01}} 
& {\textbf{73.23}} 
& {\textbf{73.95}} \\
    \cmidrule(lr){2-21}

    & \multirow{8}{*}{\makecell{API-\\Java}} 
      & No RAG & -- & -- & 35.22 & 70.11 & 83.93 & 83.81 & 41.77 & 69.40 & 63.64 & 62.83 & 43.17 & 73.81 & 84.35 & 84.95 & 47.27 & 73.09 & 67.33 & 67.37 \\
    & & Vanilla RAG & 36.28 & 48.87 & 31.82 & 63.42 & 79.12 & 78.83 & 36.42 & 63.24 & 56.37 & 55.61 & 41.37 & 70.52 & 81.74 & 82.68 & 46.32 & 69.82 & 63.50 & 63.59 \\
    & & GraphCoder & 29.64 & >60 & 42.87 & 70.34 & 82.58 & 82.79 & 47.82 & 70.23 & 64.27 & 63.85 & 53.73 & 79.37 & 87.65 & 88.05 & 58.93 & 78.81 & 74.02 & 74.03 \\
    & & RepoFuse & 11.13 & >60 & 37.43 & 65.48 & 80.33 & 79.93 & 42.56 & 67.33 & 59.43 & 58.88 & 42.42 & 72.89 & 84.28 & 84.81 & 46.67 & 72.11 & 65.97 & 66.13 \\
    & & RLCoder & -- & -- & 41.62 & 70.18 & 83.18 & 82.81 & 46.42 & 69.92 & 64.74 & 64.02 & 52.83 & 78.34 & 86.98 & 87.48 & 57.53 & 77.47 & 72.85 & 72.87 \\
    & & Naive \modelname & 1.06 & 6.09 & 40.27 & 69.70 & 82.90 & 82.50 & 45.47 & 69.69 & 63.64 & 62.99 & 51.65 & 77.32 & 85.34 & 87.16 & 56.18 & 77.04 & 71.48 & 71.06 \\
    & & \modelname & 1.16 & 6.19 & {45.67} & {73.35} & {85.60} & {85.19} & {51.08} & {73.28} & {68.00} & {67.16} & {54.93} & {79.58} & {87.96} & {88.63} & {59.37} & {79.05} & \textbf{74.21} & {74.20} \\
    & & {\modelname (0.6b)} 
& {1.07} 
& {2.41} 
& {\textbf{50.53}} 
& {\textbf{75.75}} 
& {\textbf{86.14}} 
& {\textbf{85.91}} 
& {\textbf{55.58}} 
& {\textbf{75.39}} 
& {\textbf{71.01}} 
& {\textbf{70.34}} 
& {\textbf{55.13}} 
& {\textbf{79.71}} 
& {\textbf{88.03}} 
& {\textbf{88.71}} 
& {\textbf{59.43}} 
& {\textbf{79.18}} 
& {{74.13}} 
& {\textbf{74.42}} \\
    \bottomrule
    \end{tabular}%
    }
    \begin{tablenotes}
        \tiny
        \item[*] \parbox[t]{0.95\textwidth}{RAG Pipeline Time covers end-to-end indexing and retrieval for baselines, and ripgrep generation plus retrieval for \modelname. A detailed introduction of the \modelname (0.6B) distilled variant is provided in Sec.~\ref{sec:Discussion}.}
    \end{tablenotes}
    \end{threeparttable}
\end{table*}

\subsection{Ablation Study: Dissecting the Improvement}

To quantify the contribution of each component, we conduct an ablation study on the CrossCodeEval dataset (Table~\ref{tab:ablation_full}). We compare four variants, and the results demonstrate that structure-aware deduplication contributes more substantially to performance gains:

\stitle{\modelname\ (Naive)}: 
    The baseline configuration. Due to the absence of deduplication and weighted re-ranking, its performance is constrained by redundancy and noise.

\stitle{\modelname\ w/o Dedup}: 
    Replaces the re-ranking strategy from Jaccard to BM25 exclusively (without deduplication). The results indicate relatively marginal performance gains; the EM on Python DeepSeek-V3.2-EXP increased only from 38.61\% to 39.12\%, a rise of 0.51\%. This suggests that while BM25 optimizes ranking weights, solely refining the ranking algorithm struggles to break the bottleneck of insufficient information density when the context window is occupied by a substantial volume of repetitive code fragments.
    \add{To further analyze whether the improvement brought by BM25 is statistically meaningful rather than random fluctuation, we conduct paired significance tests for all comparisons between BM25 and Jaccard using McNemar's test. The results consistently satisfy the conventional statistical significance criterion ($p<0.05$), confirming that the proposed modification is effective rather than incidental.}

\stitle{\modelname\ w/o BM25}: 
    This variant retains the original Jaccard ranking while adding the deduplication module. It achieves a substantial performance improvement, reaching an EM of 41.93\% on Python DeepSeek-V3.2-EXP—an increase of 3.32\% over Naive \modelname. This gain is considerably larger than that obtained by merely replacing the ranking algorithm.

\stitle{\modelname\ (Full)}: 
    The complete configuration. This combination achieves the optimal performance (42.29\%), demonstrating that once the deduplication mechanism secures information breadth, the precise ranking of BM25 further optimizes information precision. The two exhibit a favorable orthogonal complementarity.

\begin{table*}[htbp]
    \centering
    \caption{Ablation Study: Impact of Re-ranking and De-duplication on CrossCodeEval. }
    \label{tab:ablation_full}
    \resizebox{\textwidth}{!}{%
    \setlength{\tabcolsep}{2pt}
    \begin{tabular}{@{}llc|cccccccc|cccccccc@{}}
    \toprule
     & & & \multicolumn{8}{c|}{\textbf{DeepSeek-V3.2-EXP}} & \multicolumn{8}{c}{\textbf{Qwen3-Coder-Plus}} \\
    \cmidrule(lr){4-11} \cmidrule(lr){12-19}
     & & \textbf{}
       & \multicolumn{4}{c}{Code} & \multicolumn{4}{c|}{Identifier}
       & \multicolumn{4}{c}{Code} & \multicolumn{4}{c}{Identifier} \\
    \cmidrule(lr){4-7} \cmidrule(lr){8-11} \cmidrule(lr){12-15} \cmidrule(lr){16-19}
     \textbf{Lang} & \textbf{Method (Component)} & Retrieval Time(s)
     & EM & ES & Recall & F1 & EM & ES & Recall & F1
     & EM & ES & Recall & F1 & EM & ES & Recall & F1 \\
    \midrule

    \multirow{4}{*}{\textbf{Python}} 
      & \modelname\ (Naive)
      & 0.0186
      & 38.61 & 77.54 & 89.08 & 87.50 & 48.33 & 78.67 & 74.59 & 72.33
      & 40.79 & 79.82 & 88.96 & 88.52 & 51.52 & 80.97 & 75.96 & 74.85 \\

      & \modelname\ w/o Dedup
      & 0.0189
      & 39.12 & 77.96 & 89.42 & 87.85 & 49.21 & 79.32 & 75.47 & 73.02
      & 41.35 & 79.94 & 89.13 & 89.03 & 52.12 & 81.20 & 76.24 & 75.39 \\

      & \modelname\ w/o BM25
      & 0.0192
      & 41.93 & 79.05 & 89.49 & 88.04 & 51.83 & 80.12 & 76.19 & 74.93
      & 43.29 & 80.71 & 89.42 & 89.23 & 54.38 & 82.13 & 77.21 & 76.84 \\

      & \textbf{\modelname\ (Full)}
      & 0.0197
      & \textbf{42.29} & \textbf{79.66} & \textbf{89.53} & \textbf{88.50} & \textbf{52.76} & \textbf{80.76} & \textbf{76.84} & \textbf{75.15}
      & \textbf{44.62} & \textbf{81.32} & \textbf{89.58} & \textbf{89.45} & \textbf{55.87} & \textbf{82.62} & \textbf{77.90} & \textbf{77.29} \\

    \midrule

    \multirow{4}{*}{\textbf{Java}} 
      & \modelname\ (Naive)
      & 0.0173
      & 41.70 & 78.93 & 89.93 & 89.33 & 52.17 & 79.95 & 76.62 & 75.67
      & 41.42 & 78.31 & 87.92 & 88.67 & 51.90 & 79.02 & 74.69 & 74.77 \\

      & \modelname\ w/o Dedup
      & 0.0175
      & 42.17 & 79.35 & 89.97 & 89.42 & 52.88 & 80.02 & 76.75 & 75.81
      & 41.92 & 78.60 & 88.16 & 88.84 & 52.63 & 79.80 & 74.91 & 75.03 \\

      & \modelname\ w/o BM25
      & 0.0179
      & 42.87 & 79.92 & 89.69 & 89.77 & 53.34 & 80.62 & 77.10 & 76.31
      & 43.89 & 79.03 & 88.57 & 89.03 & 53.76 & 80.42 & 76.19 & 76.03 \\

      & \textbf{\modelname\ (Full)}
      & 0.0181
      & \textbf{43.15} & \textbf{80.07} & \textbf{89.78} & \textbf{89.83} & \textbf{53.81} & \textbf{80.88} & \textbf{77.41} & \textbf{76.57}
      & \textbf{44.09} & \textbf{79.95} & \textbf{88.98} & \textbf{89.29} & \textbf{54.09} & \textbf{80.69} & \textbf{76.83} & \textbf{76.57} \\

    \bottomrule
    \end{tabular}%
    }
\end{table*}

\subsection{Generalization of Ripgrep Command Generation}
To verify that the performance gains achieved by our framework stem from genuine model agnosticism rather than parameter biases of specific models, we evaluated the generalization capability of the ripgrep command generation module on the CrossCodeEval dataset. As shown in Table~\ref{tab:generalization}, we deployed DeepSeek-V3.2-EXP and Qwen3-Coder-Plus as instruction generators. Results show that the choice of instruction generator has little effect on downstream code completion performance. This suggests that the ability to autonomously generate code retrieval instructions is not unique to a specific model but a general capability of modern LLMs. This ensures the stability of our framework across diverse LLM architectures.

\begin{table*}[t]
    \centering
    \caption{Generalization Study of Ripgrep Command Generation across Different Instruction Generators and Code Completion Backbones on CrossCodeEval.}
    \label{tab:generalization}
    \resizebox{\textwidth}{!}{%
        \setlength{\tabcolsep}{3.5pt} 
        \begin{tabular}{ll| cccccccc| cccccccc}
            \toprule
            
            &&\multicolumn{8}{c|}{\textbf{DeepSeek-V3.2-EXP}} & 
            \multicolumn{8}{c}{\textbf{Qwen3-Coder-Plus}} \\

            \cmidrule(lr){3-10} \cmidrule(lr){11-18}

            & & \multicolumn{4}{c}{{Code}} & \multicolumn{4}{c|}{{Identifier}} & \multicolumn{4}{c}{{Code}} & \multicolumn{4}{c}{{Identifier}} \\
            \cmidrule(lr){3-6} \cmidrule(lr){7-10} \cmidrule(lr){11-14} \cmidrule(lr){15-18}

            \textbf{Lang} &\textbf{Method}& EM & ES & Recall & F1 & EM & ES & Recall & F1 & EM & ES & Recall & F1 & EM & ES & Recall & F1 \\
            \midrule

            \multirow{2}{*}{\textbf{Python}} 
            & \modelname\ w/ DeepSeek 
            & \textbf{42.29} & \textbf{79.66} & \textbf{89.53} & \textbf{88.50} & \textbf{52.76} & \textbf{80.76} & \textbf{76.84} & \textbf{75.15} 
            & \textbf{44.62} & \textbf{81.32} & \textbf{89.58} & \textbf{89.45} & \textbf{55.87} & \textbf{82.62} & \textbf{77.90} & \textbf{77.29} \\
            
            & \modelname\ w/ Qwen 
            & 41.03 & 79.03 & 89.15 & 88.38 & 52.20 & 80.20 & 75.89 & 74.69 
            & 44.21 & 80.96 & 89.05 & 89.07 & 54.91 & 82.07 & 76.70 & 76.45 \\
            
            \addlinespace 
            \midrule
            
            \multirow{2}{*}{\textbf{Java}} 
            & \modelname\ w/ DeepSeek 
            & \textbf{43.15} & \textbf{80.07} & \textbf{89.78} & \textbf{89.83} & \textbf{53.81} & \textbf{80.88} & \textbf{77.41} & \textbf{76.57} 
            & \textbf{44.09} & \textbf{79.95} & \textbf{88.98} & \textbf{89.29} & \textbf{54.09} & \textbf{80.69} & \textbf{76.83} & \textbf{76.57} \\
            
            & \modelname\ w/ Qwen 
            & 42.57 & 79.30 & 89.67 & 89.38 & 53.09 & 80.39 & 77.25 & 75.93 
            & 43.65 & 79.14 & 88.31 & 89.05 & 53.76 & 79.64 & 75.47 & 75.24 \\
            
            \bottomrule
        \end{tabular}%
    }
\end{table*}

\subsection{Hyperparameter Sensitivity Analysis}
\begin{figure}[htbp]
  \centering

  \begin{subfigure}[t]{0.48\textwidth}
    \centering
    \includegraphics[width=\linewidth]{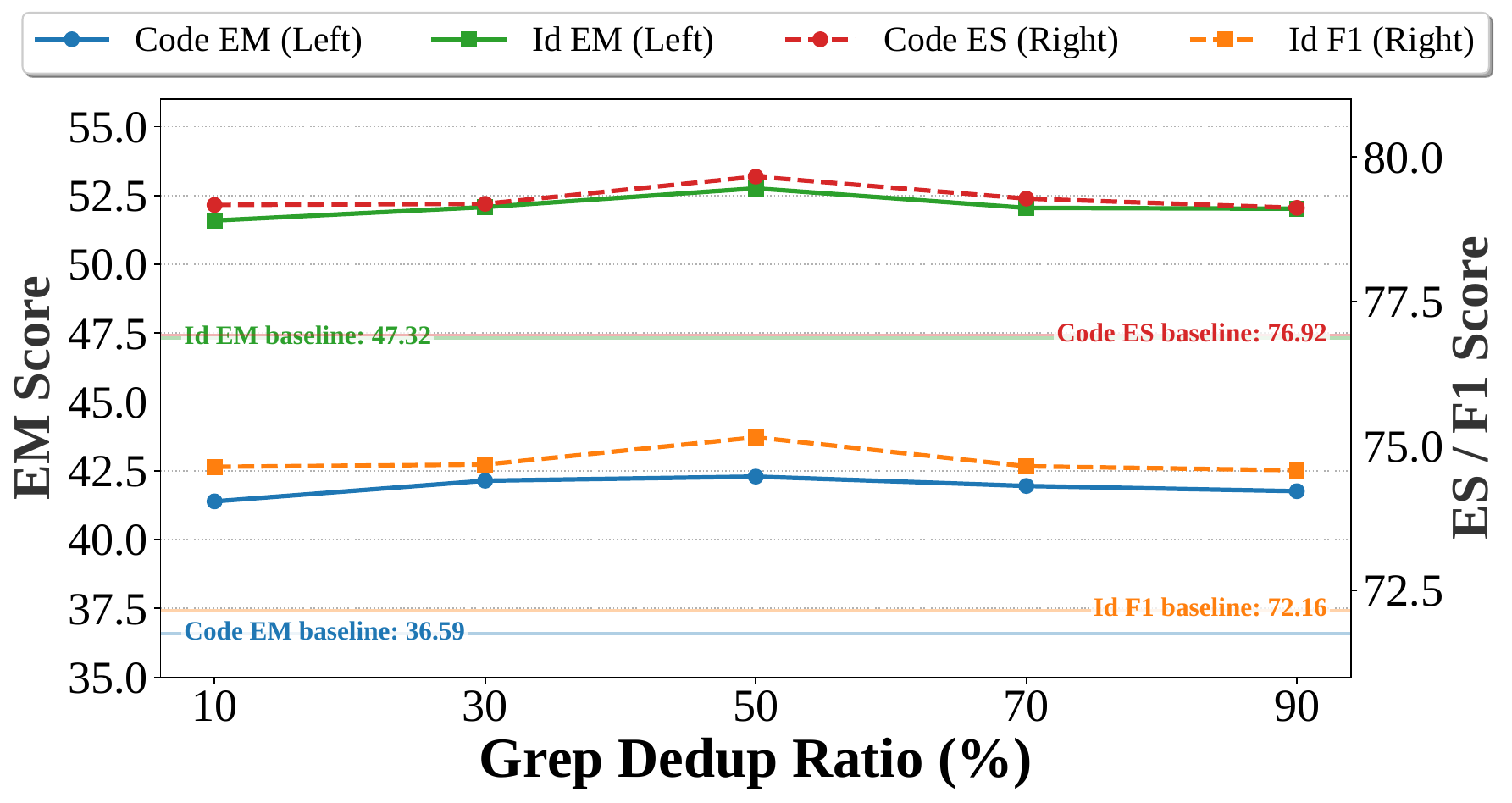}
    \caption{CrossCodeEval-Python}
  \end{subfigure}\hfill
  \begin{subfigure}[t]{0.48\textwidth}
    \centering
    \includegraphics[width=\linewidth]{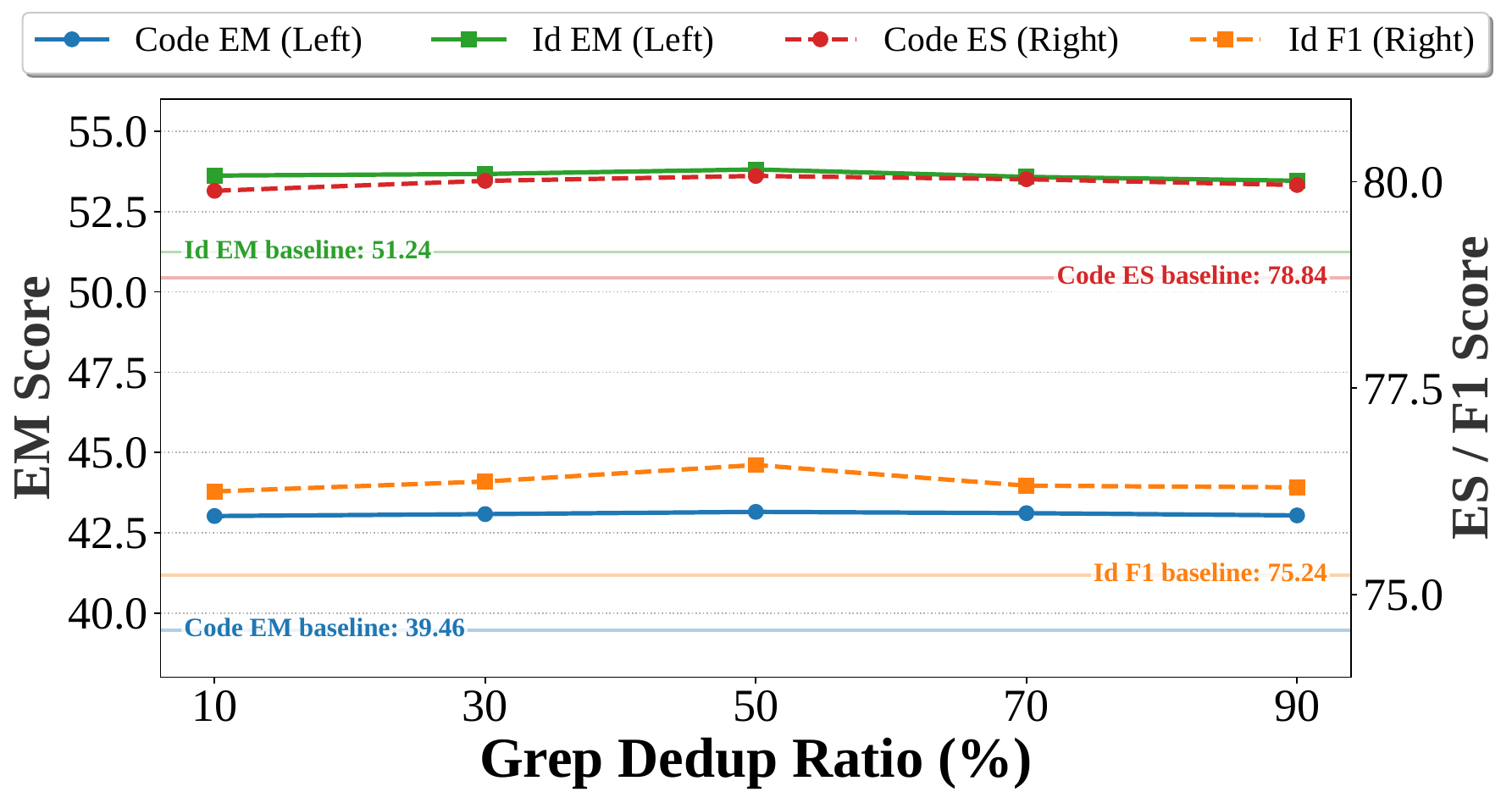}
    \caption{CrossCodeEval-Java}
  \end{subfigure}

  \vspace{0.8em}

  \begin{subfigure}[t]{0.48\textwidth}
    \centering
    \includegraphics[width=\linewidth]{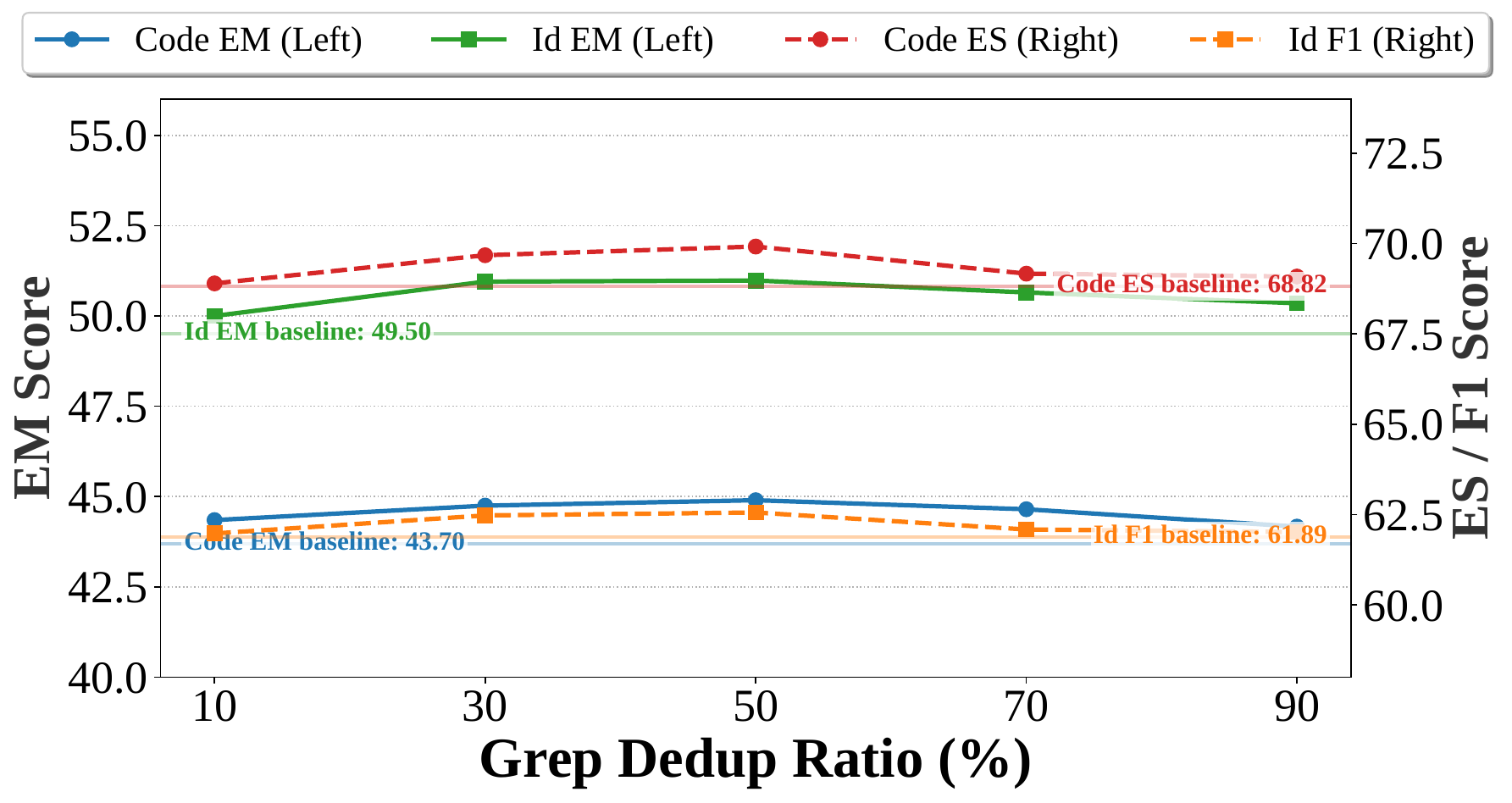}
    \caption{RepoEval\_Updated-Python}
  \end{subfigure}\hfill
  \begin{subfigure}[t]{0.48\textwidth}
    \centering
    \includegraphics[width=\linewidth]{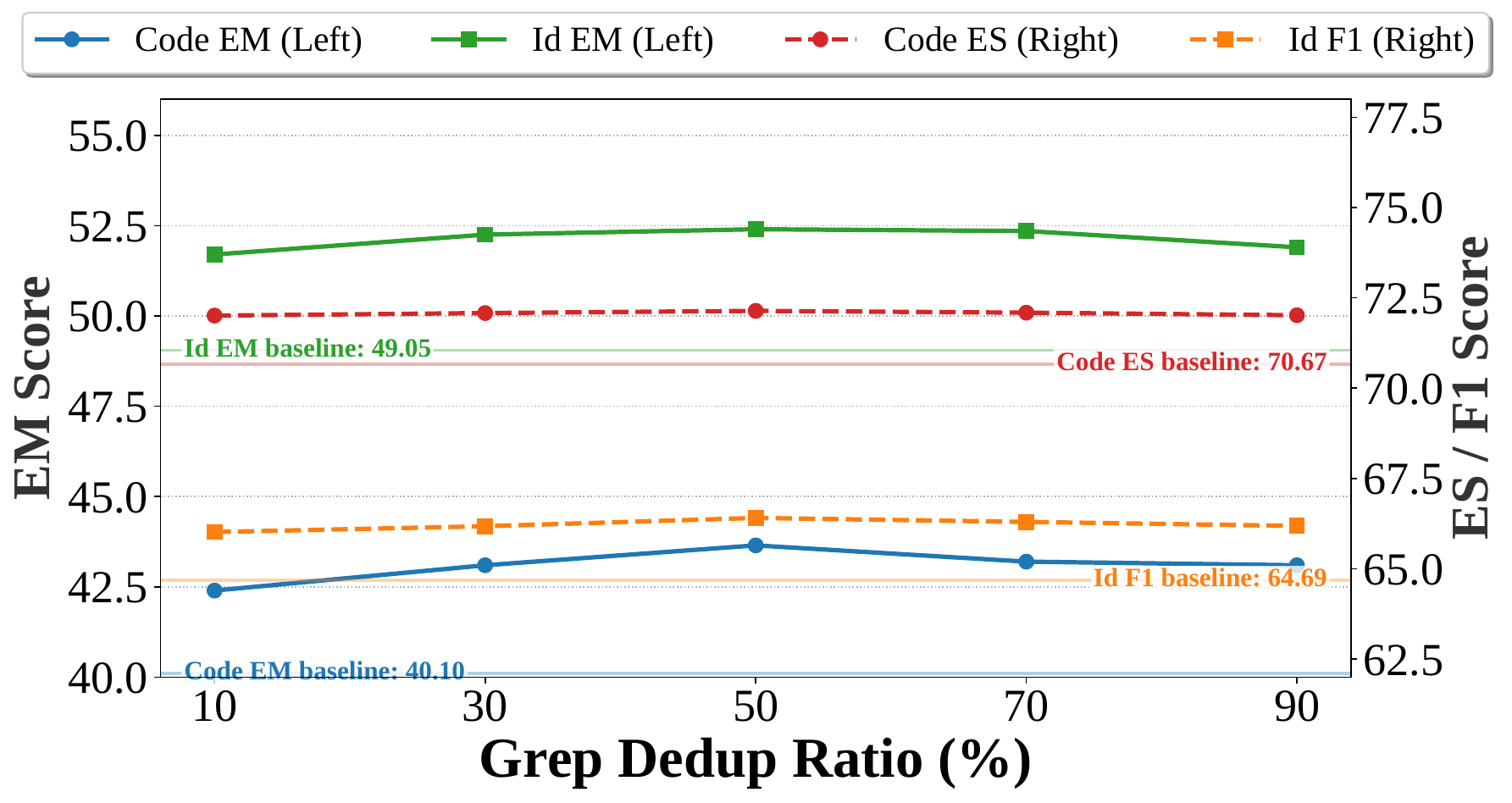}
    \caption{RepoEval\_Updated-Java}
  \end{subfigure}

  \caption{Sensitivity analysis of the de-duplication candidate pool size ($N$). Horizontal dashed lines denote the strongest baseline for each metric in Table~\ref{tab:merged-performance}. Across both CrossCodeEval and RepoEval\_Updated, \modelname{} remains robust under $N=10\%$--$90\%$. Although performance generally peaks around $N=50\%$, all tested settings consistently outperform the strongest baseline.}
  \label{fig:hyperparam_all}
\end{figure}

\delete{On the large-scale RepoEval\_Updated dataset, we investigated a critical hyperparameter $N$ within the post-processing stage.}
\add{We further analyze the sensitivity of the hyperparameter $N$ within the post-processing stage on both CrossCodeEval and RepoEval\_Updated.}
This parameter denotes the percentage of candidate fragments (Top-N\%) selected from the BM25 ranked list prior to the execution of the deduplication operation. 
We comprehensively evaluated the trends of four core metrics (code match EM/ES, identifier match EM/F1) as the value of $N$ varies from 10\% to 90\%, as illustrated in Figure \ref{fig:hyperparam_all}.

\delete{The experimental results exhibit a highly consistent inverted U-shaped pattern, indicating that the selection of $N$ is pivotal for constructing high-quality context.}

When $N < 50\%$, performance drops markedly. Due to severe redundancy in large repositories, the head of the BM25 list is often dominated by semantically identical blocks from different locations. Consequently, after performing deduplication, the effective Top list may collapse to only 1--2 blocks, leaving the LLM context window underutilized.

At $N = 50\%$, the model consistently achieves peak performance on both Python and Java across all metrics. This indicates that $N = 50\%$ provides a robust balance between candidate coverage and information density. We therefore adopt $N = 50\%$ as the default setting in \modelname.
\add{More importantly, the overall variation across different values of $N$ is relatively small: all settings from 10\% to 90\% consistently outperform the strongest baseline on both datasets, indicating that the gains of \modelname do not rely on a narrowly tuned hyperparameter choice and are robust to the selection of $N$.}

When $N > 50\%$, performance plateaus or slightly degrades, as low-relevance blocks from the tail of the BM25 ranking contribute little to Top-K selection and may introduce additional noise.

\section{Discussion}
\label{sec:Discussion}

\textbf{Mitigating LLM Overhead for Retrieval Command Generation}
Although the physical retrieval cost of \texttt{ripgrep} is minimal (on the order of milliseconds), generating retrieval commands with a general-purpose LLM still incurs additional inference latency due to the \textbf{model scale} and \textbf{output token length}. To address this issue, we optimize the process from two complementary perspectives: \textbf{output constraints} and \textbf{knowledge distillation}.

Regarding output length, we observe that \texttt{grep} commands exhibit highly templated characteristics, such as fixed parameters and rigid formats. Accordingly, when fine-tuning Qwen3-0.6B, the model is tasked only with predicting the core retrieval keywords, while the remaining command structure is directly instantiated from a static template. This design substantially reduces the number of generated tokens and thus lowers inference latency. For distillation, since the quality of the teacher model directly determines the performance upper bound of supervised fine-tuning~\cite{zhang2025openmmreasoner}, we employ \texttt{claude-opus-4-5-20251101}, a model with strong code generation capabilities, as the teacher to construct high-quality training data.

\add{Results on all evaluation datasets are reported in Table~\ref{tab:merged-performance}.}
On the line-level Python and Java tasks of the RepoEval\_Updated dataset, the fine-tuned 0.6B model surpasses substantially larger general-purpose models in completion quality, achieving a dual optimization of performance and computational cost.

The RAG pipeline time reported in the table includes the end-to-end cost of indexing and retrieval for baseline methods, while for \modelname\ it refers to the combined time of \texttt{ripgrep} command generation and retrieval execution. The \modelname\ (DeepSeek-V3.2-EXP) variant is excluded from this comparison due to additional network latency introduced by API-based \texttt{ripgrep} command invocation. We observe that \modelname\ (0.6B Distilled) exhibits substantially lower RAG pipeline time than graph-based baseline methods. Moreover, the time required for \texttt{ripgrep} command generation has constant time complexity ($O(1)$), as it depends only on the local context of the current editing window and is independent of repository size. In contrast, graph-based approaches typically incur $O(N)$ or higher time overhead as the repository scale grows. Consequently, our method offers superior scalability and deployment practicality for large-scale, frequently evolving industrial code repositories where index maintenance is often a bottleneck.

\textbf{Potential Data Contamination.} 
A primary potential threat lies in the possibility that the pre-training corpora of LLMs may encompass portions of the evaluation benchmarks (CrossCodeEval/RepoEval\_Updated), thereby introducing {memorization bias} into the assessment results. Although we cannot entirely rule out data overlap, the core conclusions of this study are premised on {relative performance gains}, rather than absolute scores. All baseline methods and our proposed approach utilize identical backbone models. The significant improvements observed in \modelname\ relative to baselines provide compelling evidence that the performance gains stem from more precise context retrieval, rather than knowledge leakage within the model parameters.

\add{
\textbf{Comparison with Coding Agents.}
Although coding agents are designed for interactive repository-level software engineering rather than standalone code completion, we further compare GrepRAG with Claude Code under the same DeepSeek-V3.2 backbone. We evaluate on RepoEval\_Updated (line-level and API-level) and CrossCodeEval. Due to the high time cost of running Claude Code on full datasets, we randomly sample from each full evaluation configuration. Specifically, RepoEval\_Updated uses 372 samples for each task-language configuration, while CrossCodeEval uses 446 Python samples and 439 Java samples. These sample sizes satisfy a 95\% confidence level with at most 5\% margin of error.
}

\add{
For the agent setup, we use Claude Code v2.1.98 with a standardized prompt: ``Complete the next line of \texttt{<file\_name>} and only output the completion result.'', where \texttt{<file\_name>} denotes the target file. The agent is given access to the full repository during inference.
}

\begin{table*}[t]
\centering
\small
\setlength{\tabcolsep}{2.5pt}
\caption{Comparison between GrepRAG and Claude Code on randomly sampled sets.}
\label{tab:agent_detailed_metrics}
\begin{tabular}{llllrrrrrrrr}
\toprule
\textbf{Dataset} & \textbf{Task} & \textbf{Lang} & \textbf{Method}
& \multicolumn{4}{c}{\textbf{Code}} & \multicolumn{4}{c}{\textbf{Identifier}} \\
\cmidrule(lr){5-8} \cmidrule(lr){9-12}
& & & & EM & ES & Recall & F1 & EM & ES & Recall & F1 \\
\midrule
\multirow{8}{*}{RepoEval\_Updated} & \multirow{4}{*}{Line} & \multirow{2}{*}{Python} & GrepRAG & \textbf{45.16} & \textbf{68.77} & \textbf{84.84} & \textbf{82.60} & \textbf{49.73} & \textbf{70.18} & \textbf{66.64} & \textbf{64.02} \\
 & & & Claude Code & 43.28 & 60.05 & 78.56 & 75.05 & 45.70 & 60.89 & 58.14 & 55.78 \\
 \cmidrule(lr){3-12}
 & & \multirow{2}{*}{Java} & GrepRAG & \textbf{45.43} & \textbf{72.70} & \textbf{84.69} & \textbf{83.94} & \textbf{55.65} & \textbf{75.04} & \textbf{72.10} & \textbf{70.06} \\
 & & & Claude Code & 36.02 & 57.15 & 73.40 & 71.13 & 44.62 & 59.82 & 57.49 & 54.78 \\
\cmidrule(lr){2-12}
 & \multirow{4}{*}{API} & \multirow{2}{*}{Python} & GrepRAG & \textbf{44.89} & \textbf{72.55} & \textbf{86.26} & \textbf{86.54} & \textbf{49.19} & \textbf{73.22} & \textbf{67.81} & \textbf{68.46} \\
 & & & Claude Code & 37.63 & 54.55 & 81.18 & 76.62 & 40.05 & 56.31 & 53.23 & 50.47 \\
 \cmidrule(lr){3-12}
 & & \multirow{2}{*}{Java} & GrepRAG & \textbf{45.97} & \textbf{72.20} & \textbf{84.42} & \textbf{84.44} & \textbf{51.08} & \textbf{72.23} & \textbf{66.16} & \textbf{65.69} \\
 & & & Claude Code & 36.83 & 54.05 & 77.29 & 73.30 & 38.98 & 54.67 & 51.18 & 47.96 \\
\midrule
\multirow{4}{*}{CrossCodeEval} & \multirow{4}{*}{Line} & \multirow{2}{*}{Python} & GrepRAG & \textbf{43.50} & \textbf{81.15} & \textbf{89.63} & \textbf{89.10} & \textbf{52.47} & \textbf{82.22} & \textbf{77.58} & \textbf{76.53} \\
 & & & Claude Code & 37.44 & 55.20 & 86.28 & 76.55 & 40.13 & 56.42 & 60.06 & 50.17 \\
 \cmidrule(lr){3-12}
 & & \multirow{2}{*}{Java} & GrepRAG & \textbf{44.65} & \textbf{80.13} & \textbf{89.11} & \textbf{89.21} & \textbf{53.76} & \textbf{81.24} & \textbf{77.54} & \textbf{76.96} \\
 & & & Claude Code & 28.25 & 50.97 & 80.76 & 74.21 & 31.44 & 51.68 & 51.02 & 44.03 \\
\bottomrule
\end{tabular}
\end{table*}

\add{
Overall, Table~\ref{tab:agent_detailed_metrics} shows that GrepRAG consistently achieves higher code-level and identifier-level scores across RepoEval\_Updated and CrossCodeEval. Moreover, Claude Code requires 97.97 s per instance on average, making it substantially slower than GrepRAG. These results suggest that Claude Code + DeepSeek-V3.2 is less effective than GrepRAG under the same backbone model.
}

\section{Related Work}
\subsection{Large Language Models for Code}

The rapid evolution of LLMs has profoundly reshaped AI research, advancing natural language and multi-modal understanding\add{~\cite{bai2025qwen2,yang2025longvt,yao2025rigorous,wang2026deepresearcheval}} while significantly boosting automated code completion and other software engineering tasks~\cite{zan2022large,xu2025prioritizing,vaswani2017attention,wu2025llmapphub,nejjar2025llms,wang2025exploracoder,xu2025revisiting,wang2025inspectcoder}. Contingent upon the accessibility of model weights, existing technical paradigms are primarily categorized into two distinct classes: proprietary closed-source and open-source. Within this landscape, proprietary models\cite{achiam2023gpt,team2023gemini}, such as the GPT-5.2 series, Gemini3, and Claude Opus 4.5, are generally regarded as performance benchmarks for evaluating code generation capabilities. Simultaneously, the open-source community offers a plethora of robust alternatives, encompassing both general-purpose foundation models such as DeepSeek-V3\cite{guo2024deepseek,liu2024deepseek,liu2024deepseekv2} and Qwen 3\cite{hui2024qwen2,ahmed2025qwen,yang2025qwen2,yang2025qwen3}, as well as domain-specific models explicitly optimized for programming languages, including Code Llama\cite{roziere2023code} and StarCoder\cite{li2023starcoder,lozhkov2024starcoder}. 

\subsection{Repository-Level Code Completion}

While the aforementioned models excel in handling intra-file logic~\cite{lozhkov2024starcoder}, they struggle in repository-level scenarios. Due to the highly modular nature of code logic, critical class definitions, function interfaces, and global constants are typically dispersed across disparate files. Consequently, models constrained by limited context windows fail to capture global semantics~\cite{liu2023lost,shi2023large,ni-etal-2024-l2ceval}. To address this challenge, RAG techniques have been introduced, aiming to retrieve relevant cross-file context from the global codebase to augment the generation process~\cite{shrivastava2023repofusion,shrivastava2023repository,tan2024prompt,ding2022cocomic,xiao2024c}. Existing technical paradigms have evolved primarily along three distinct directions.

Early research predominantly adopted {similarity-based retrieval paradigms}, utilizing semantic or lexical matching to locate reference information~\cite{hashimoto2018retrieve,li2022coderetriever,lu2022reacc}. For instance, AceCoder~\cite{li2023acecoder} and APICoder~\cite{zan2022language} retrieve similar code snippets and API documentation, respectively, whereas RepoCoder~\cite{zhang2023repocoder} dynamically expands query semantics through an iterative Generate-then-Retrieve loop. Although these methods extend the context window, their reliance on BM25 or vector similarity makes it difficult to capture the intrinsic logical dependencies of code.
In response, {structure-aware methods} incorporate static analysis techniques, attempting to explicitly model the topological relationships of code~\cite{guo2020graphcodebert,allamanis2017learning}. Works such as GraphCoder~\cite{liu2024graphcoder}, RepoHyper~\cite{phan2024repohyper}, and Cocomic~\cite{ding2022cocomic} construct code context graphs or repository-wide dependency graphs to acquire structured information; RepoFuse~\cite{liang2024repofuse} further integrates noise filtering mechanisms upon this foundation. However, complex graph construction and traversal processes inevitably introduce high computational latency.
Recent research has shifted towards {strategy optimization and alignment}, striving to bridge the objective gap between retrieval and generation tasks. RLCoder~\cite{wang2024rlcoder} employs reinforcement learning for end-to-end fine-tuning of the retriever, utilizing generation probability as the optimization objective. Similarly, AlignCoder\add{~\cite{jiang2025aligncoder}} enhances query semantics by generating candidate completions and trains a dedicated retriever using feedback signals, thereby enabling the retrieval strategy to dynamically adapt to the inference requirements of downstream models.
\section{Conclusion}
This paper demonstrates the effectiveness of lightweight lexical retrieval for code completion. Our experiments show that Naive \modelname can achieve competitive performance by capturing explicit lexical dependencies, but suffers under noisy and fragmented contexts. To overcome these limitations, we propose \modelname, combining identifier-weighted re-ranking with structural deduplication. Experiments on CrossCodeEval and RepoEval\_Updated demonstrate consistent improvements over strong baselines, establishing a new SOTA. Future work will investigate adaptive routing to better support implicit dependency scenarios.
\section{Data Availability}
\label{sec:data_availability}
To ensure the reproducibility of our findings, we have made the source code, pre-processed datasets, and evaluation scripts available at \url{https://github.com/ZJU-ACES-ISE/greprag}.
\begin{acks}
This work was supported in part by the National Natural Science Foundation of China under Grant No. 62402433, in part by the Natural Science Foundation of Zhejiang Province under Grant No. LZ25F020002, in part by the Zhejiang Pioneer (Jianbing) Project under Grant No. 2025C01198(SD2), in part by the Scientific Research Foundation of Hangzhou City University under Grant No. J-202608, and in part by Zhejiang Key Laboratory Project under Grant No. 2024E10001.
\end{acks}

\bibliographystyle{ACM-Reference-Format}
\bibliography{ref}

\appendix


\end{document}